\documentclass[prr, reprint, twocolumn, amsmath,amsfonts,amssymb]{revtex4-1}
\usepackage{color}
\usepackage{bm}
\usepackage{graphicx}
\usepackage{array,longtable}
\usepackage{threeparttable}
\usepackage{here}

\usepackage[T1]{fontenc}       % Use modern font encodings
\usepackage{graphicx}
\usepackage{color}
\usepackage{bm}
\usepackage{threeparttable}
\usepackage[mathscr]{eucal}
\usepackage{ulem}
\usepackage{picture}  
\begin{document}
\title{Spin-projection for quantum computation: A low-depth approach to
strong correlation}
\author{Takashi Tsuchimochi}
\email{tsuchimochi@gmail.com}
\author{Yuto Mori}
\affiliation{Graduate School of System Informatics, Kobe University, Kobe 657-8501, Japan}
\author{Seiichiro L. Ten-no}
\affiliation{Graduate School of System Informatics, and Graduate School of Science, Technology, and Innovation, Kobe University, Kobe 657-8501, Japan}

\begin{abstract}
Although spin is a core property in fermionic systems, its symmetry can be easily violated in a variational simulation, especially when strong correlation plays a vital role therein. In this study, we will demonstrate that the broken spin-symmetry can be restored exactly in a quantum computer, with little overhead in circuits, while delivering additional strong correlation energy with the desired spin quantum number. The proposed scheme permits drastic reduction of a potentially large number of measurements required to ensure spin-symmetry by employing a superposition of only a few rotated quantum states. Our implementation is universal, simple, and, most importantly, straightforwardly applicable to any ansatz proposed to date.
\end{abstract}

\maketitle 
 
\section{Introduction} Recent advancements on quantum devices have created widespread interest in the development of efficient quantum algorithms. The Variational Quantum Eigensolver (VQE) is one of the most pursued approaches in the NISQ (noisy intermediate-scale quantum) era, where a trial state $|\psi({\bm\theta})\rangle$ parameterized by ${\bm\theta}$ is variationally optimized to minimize its Hamiltonian expectation value.\cite{Peruzzo14, OMalley16} Among the several candidates for $|\psi\rangle$, the unitary coupled cluster with singles and doubles (UCCSD)\cite{Kutzelnigg77,Bartlett89,Taube06} ansatz has been extensively used as an entangler in the preparation of a trial state from Hartree-Fock (HF) $|\Phi\rangle$, as shown by the following equation:
\begin{equation}
\label{eq:UCCSD}
|\psi_{\rm UCCSD}\rangle = e^{\hat {\cal T}_1 + \hat {\cal T}_2} |\Phi\rangle,\;\;\;\hat{\cal T}_k = \sum_{ab\cdots,ij\cdots}  t_{ij\cdots}^{ab\cdots} \hat \tau_{ij\cdots}^{ab\cdots}
\end{equation}
Here, $t_{ij\cdots}^{ab\cdots}$ and $\hat \tau_{ij\cdots}^{ab\cdots}= 	\left(a^\dag_{a} a^\dag_{b} \cdots a_{j} a_{i} -h.c.\right)$ are real parameters and anti-hermitian pairs of the $k$th excitation and de-excitation operators in a spin-orbital basis. In this work, we use convention for the orbital indices: $i,j$ for the occupied orbitals of $|\Phi_{\rm HF}\rangle$, $a,b$ for the virtual orbitals, and $p,q,r,s$ for the general orbitals. To make the unitary exponential operator programmable on a quantum device, Trotterization is required in practice for non-commutative exponents, i.e., $e^{\hat {\cal T}_1+\hat {\cal T}_2} \approx \left(\prod_{ai} e^{t_{i} ^{a}\hat \tau_i^a/\mu }\prod_{abij}e^{t_{ij}^{ab}\hat\tau_{ij}^{ab}/\mu  }\right)^\mu$.\cite{Barkoutsos18,Moll18,Romero19} However, applying UCCSD to strongly correlated systems often triggers large $t$-amplitudes to account for higher excitations, which, in turn, necessitates a large Trotter number $\mu$, thereby requiring a high-depth quantum circuit. Since Trotterization is nothing but an artifact needed to respect the ansatz Eq.~(\ref{eq:UCCSD}),  many authors have studied the efficacy of a single Trotter step ($\mu=1$).\cite{Barkoutsos18, Evangelista19, Sokolov20} More flexible ans\"atze have been also proposed, where, in the product of exponential operators, the same anti-hermitian excitations may be repeated albeit with different amplitudes.\cite{Lee19,Grimsley19}

To our knowledge, these recent studies on UCC and its variants have vastly neglected the spin properties of the obtained solutions. Thus, their discussions remained somewhat ambiguous. We argue that it is extremely important to monitor $\langle\hat S^2\rangle$ because, even with a spin-restricted HF (RHF) reference, UCC amplitudes can spontaneously violate spin-symmetry, thereby variationally lowering its energy as opposed to traditional CC. This often happens especially in strongly correlated systems, such as bond dissociations, as will be demonstrated below. Such broken-symmetry (BS) solutions are not physical for non-relativistic Hamiltonians. Moreover, the problems caused by spin-contamination are well known and documented.\cite{Handy85, Andrews91, Tsuchimochi10B} This so-called {\it symmetry dilemma} poses a challenge in quantum computation. As will be shown, except for singles,  using spin-free generators {\it cannot} fix this problem because each term is not necessarily commutative. Thus, such “spin-adapted” (SA) methods \cite{Paldus77, Scuseria87} still incur Trotterization for large $t_{ij}^{ab}$. Some of the previous studies on the matter have suggested the use of a constrained approach,\cite{Andrews91, McClean16, Yen19, Ryabinkin19} wherein the Hamiltonian is augmented with a penalty term $\lambda\left(\hat S^2 - s(s+1)\right)^2$, where $\lambda \rightarrow \infty$ allows the elimination of spin-contamination. However,  whereas $\langle \hat S^2\rangle$ can be evaluated with the same effort as the energy by measuring $O(n^4)$ terms in the operator, the measurement of the expectation value of $\hat S^4$ has a steep $O(n^8)$ scaling, where $n$ is the number of qubits (i.e., spin orbitals), compared to the $O(n^4)$ scaling of the bare Hamiltonian. Ryabinkin et al. have attempted to avoid this prohibitive scaling by constraining the average spin $\langle \hat S^2\rangle$.\cite{Ryabinkin19} Nevertheless, the constrained approach does not seem particularly suited to spin as the Trotterized UCCSD ansatz is not flexible enough, provoking the symmetry dilemma. The resulting energy can become unreliable depending on the choice of $\lambda$, if the ansatz is not flexible enough to deal with strong correlation. 

In this work, we propose an alternative way of preserving the spin of an {\it arbitrary} quantum state in a numerically exact, but much less costly, manner, while simultaneously treating strong correlation. 
 To this end, we will let a quantum state break its spin symmetry, which is subsequently recovered by means of symmetry projection. By writing a spin-adapted state as a superposition of rotated broken-symmetry states, the quantum circuit we develop here fully appreciates the advantage of VQE by replacing a potentially long quantum circuit to describe entanglements with many measurements with short circuits.

 We organize the present work as follows. In Section \ref{sec:SAUCCSD}, we introduce spin-adapted UCCSD for quantum computing, and in Section \ref{sec:SCUCCSD}, we demonstrate the practical difficulty in the spin-constrained approach to handle the Lagrange multiplier $\lambda$. Section \ref{sec:SPUCCSD} then introduces the spin-projection technique for VQE as well as the required quantum circuit. Section \ref{sec:results} provides some illustrative calculations that are challenging for standard UCC. Finally, we draw our conclusions in Section \ref{sec:conclustions}.
 
\section{Theory}

\subsection{Spin-adapted UCCSD}\label{sec:SAUCCSD}
To circumvent the variational collapse via spin-symmetry breaking in UCCSD, one can introduce spin-adapted (SA)-UCCSD, where the $\hat {\cal T}_1$ and $\hat {\cal T}_2$ operators are constructed with unitary group generators $\hat E_i^a = a^\dag_a a_i + a^\dag_{\bar a} a_{\bar i}$, with a bar indicating the $\beta$ spin ($\alpha$ without a bar). That is, we use the same amplitude for spin-complement operators,
\begin{subequations}\begin{align}
t_i^a \left(\hat E_i^a - \hat E_a^i\right) &= t_i^a \left(\hat \tau_i^a + \hat \tau_{\bar i}^{\bar a}\right)\\
t_{ij}^{ab} \left(\hat E_{i}^{a}\hat E_j^b - \hat E_b^j \hat E_a^i\right) &= t_{ij}^{ab} \left( \hat \tau_{ij}^{ab} + \hat \tau_{i\bar j}^{a \bar b} + \hat \tau_{\bar i j}^{\bar a b} + \hat \tau_{\bar i \bar j}^{\bar a \bar b}\right),
\end{align}\label{eq:UGG}\end{subequations}
or, equivalently, introduce the following restrictions,\cite{Scuseria87}
\begin{subequations}
\begin{align}
	&t_i^a = t_{\bar i}^{\bar a},\\
	&t_{ij}^{ab} = t_{\bar i \bar j}^{\bar a \bar b} = t_{i\bar j}^{a\bar b} - t_{i \bar j}^{b\bar a},\\
	&t_{i\bar j}^{a\bar b} = t_{\bar i j}^{\bar a b}. 
\end{align}\label{eq:SA-UCCSD}\end{subequations}
We note that SA-UCCSD uses exactly the same circuit as BS-UCCSD but with less number of amplitudes. Hence, SA-UCCSD does not seem advantageous in terms of circuit depth.

It can be immediately realized that SA-UCCSD still cannot in principle preserve the correct spin exactly without Trotterization. Applying, for instance,  $e^{t_{ij}^{ab} \hat \tau_{ij}^{ab}}e^{t_{ij}^{ab} \hat \tau_{i\bar j}^{a\bar b}} e^{t_{ij}^{ab} \hat \tau_{\bar ij}^{\bar ab}} e^{t_{ij}^{ab} \hat \tau_{\bar i\bar j}^{\bar a\bar b}}$ as a low order approximation to $e^{t_{ij}^{ab}(\hat E_i^a\hat E_j^b - h.c.)}$ necessarily destroys spin-symmetry. This inexactness also holds for generalized doubles, which uses general orbitals. Hence, we assume that, for example, the adaptive approach of Grimsley {\it et al.} \cite{Grimsley19} can suffer from spin-contamination (incidentally, we should also point out that the operator set employed in Ref.[\onlinecite{Grimsley19}] does not constitute spin-free operators anyway, treating $\{\hat \tau_{rs}^{pq}, \hat \tau_{\bar r\bar s}^{\bar p\bar q}\}$ and $\{\hat \tau_{r\bar s}^{p\bar q}, \hat \tau_{\bar r s}^{\bar p q}\}$ differently as opposed to the above equation). The dilemma is that the extent of spin-contamination varies by size of amplitudes; so, for strongly correlated systems where $t$ amplitudes can be significantly large to describe higher excitation effects, one should pay attention to $\langle \hat S^2\rangle$ and Trotterization may be eventually required to correctly obtain the desired spin state. 

\subsection{Spin-constrained UCCSD}\label{sec:SCUCCSD}
\begin{figure}[h!]
	\includegraphics[width=1.\linewidth]{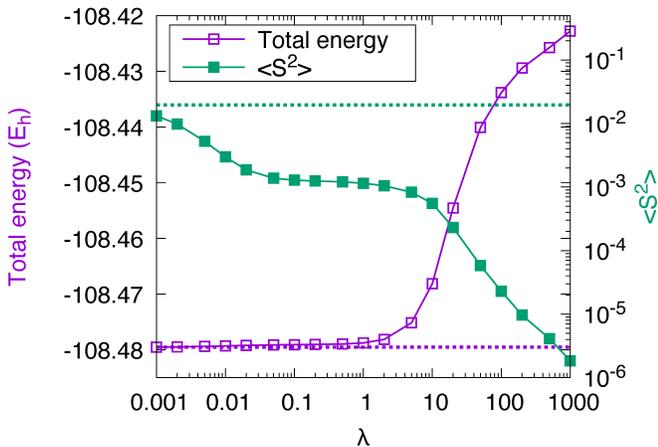}
	\caption{Spin-constrained UCCSD energy and its spin expectation value $\langle \hat S^2\rangle$ of the singlet nitrogen molecule at a fixed bond length of 2.8 {\AA} for different $\lambda$. The dashed lines indicate the values computed with $\lambda = 0$.}
	
	\label{fig:CUCCSD}
\end{figure}

A constraint on the spin $s$ can be applied to (Trotterized) UCCSD by augmenting the energy functional with a penalty term,
\begin{align}
	L[{\bm\theta}] =& \langle \psi_{\rm UCCSD}({\bm\theta}) | \hat H |\psi_{\rm UCCSD}({\bm\theta})\rangle 
	\nonumber\\ &+ \lambda \langle \psi_{\rm UCCSD}({\bm\theta})  |\left(\hat S^2 - s(s+1)\right)^2|\psi_{\rm UCCSD}({\bm\theta})\rangle.\label{eq:SCUCCSD}
\end{align}
By increasing $\lambda$, $E_{\rm UCCSD}[{\bm\theta}] \equiv \langle \psi_{\rm UCCSD}({\bm\theta}) | \hat H |\psi_{\rm UCCSD}({\bm\theta})\rangle$ becomes less spin-contaminated. However, when  $|\psi_{\rm UCCSD}\rangle$ is Trotterized, this becomes a very challenging task. Figure \ref{fig:CUCCSD} shows the total energy of Trotterized UCCSD ($\mu=1$) and its spin expectation value $\langle \hat S^2\rangle$ of the singlet nitrogen molecule at a bond length of 2.8 {\AA} by varying $\lambda$. We have used the STO-6G basis set with 1$s$ and 2$s$ orbitals frozen. While $\lambda$ is small, the energy is almost unchanged but so is $\langle \hat S^2\rangle$, suffering from spin-contamination. However, requiring the latter to be sufficiently precise with a large $\lambda$ introduces an enormous increase in the evaluated energy, which can be on the order of tens of mHartree. With such a significant energy change, it is extremely difficult to determine the appropriate $\lambda$. This test case clearly demonstrates the Trotterized UCCSD ansatz is not flexible enough to capture both the correct spin and strong correlation. To eliminate the strong dependency of energy on $\lambda$, one is required to employ large $\mu$, with which we presume Eq.~\ref{eq:SCUCCSD} would at least give the result of a similar quality to SA-UCCSD.

\subsection{Spin-Projection}\label{sec:SPUCCSD}
\subsubsection{VQE algorithm}
The spin-projection operator $\hat {\cal P}(s)$ has proven itself useful not only in removing spin contamination but also in capturing strong correlations   with its inherent high excitation effects when used with BS ans\"atze.\cite{Schlegel86,Schlegel88,Knowles88B} This is manifested from the well-known form due to L\"owdin, which is written as a product of $\hat S^2$ operators.\cite{Lowdin55B} 
From this perspective, one can expect that the role of exponential doubles $e^{t_{ij}^{ab} \hat \tau_{ij}^{ab}}$ in the spin-projected UCCSD, $|\tilde \psi_{\rm UCCSD}\rangle =\hat {\cal P} |\psi_{\rm UCCSD}\rangle$, should be to mainly describe weak dynamical correlation, with moderately small amplitudes $t_{ij}^{ab}$. Hence, the introduction of $\hat {\cal P}$ can lead to tremendous error reduction in the Trotter approximation in UCCSD.  However, despite its formal simplicity and appealing properties, the many-body nature of L\"owdin's $\hat {\cal P}$ makes it intractable, even with a quantum computer, due to the need of evaluating $\hat H\hat S^2, \hat H\hat S^4, \cdots$, which results in the exponential increase in the number of Pauli operators,\cite{Yen19} hindering its practical usage.

 Hence, we seek the possibility of the different representation of $\hat {\cal P}$, which is particularly suitable for quantum computation. It is known that a projector onto the Hilbert space of spin $s$ and spin magnetization $m$ is written in the following integral form,\cite{Percus62,Scuseria11,Jimenez12}
\begin{equation}
\label{eq:ExactP}
	\hat {\cal P}(s,m) = |s;m\rangle \langle s;m| =\int_\Omega d\Omega \; D_{mm}^{s*}(\Omega)\;  \hat {\cal U}(\Omega) 
\end{equation}
where $\Omega=(\alpha,\beta,\gamma)$ are the Euler angles, $D_{mm}^s(\Omega) = \langle s;m| \hat R(\Omega) |s;m\rangle$ is the Wigner $D$-matrix, and $\hat {\cal U}(\Omega)$ is known as the spin-rotation operators,
\begin{equation}
	\hat {\cal U}(\Omega) = e^{-i\alpha \hat S_z}e^{-i\beta \hat S_y}e^{-i\gamma \hat S_z},\label{eq:UOmega}
\end{equation}
whose resemblance to the Pauli-rotation operations is striking.  Since each of the exponents of $\hat {\cal U}(\Omega)$ is an anti-hermitian one-body operator, the many-body effect of the L\"owdin operator has been translated to a (infinite) linear combination of orbital rotations. Stated differently, a spin-projected state $|\tilde\psi({\bm\theta})\rangle = \hat {\cal P} |\psi({\bm\theta})\rangle$ with $|\psi({\bm\theta})\rangle$ being a broken-symmetry state, is understood as a superposition of broken-symmetry nonorthogonal states $ \hat {\cal U}(\Omega) |\psi({\bm\theta})\rangle$.  
For a spin-free observable $\hat O$, its expectation value is simplified as $\langle \hat {\cal P}^\dag \hat O \hat {\cal P}\rangle \equiv \langle \hat O \hat {\cal P}\rangle$. 
Because of the non-unitary operation of $\hat {\cal P}$,  $|\psi\rangle$ does not preserve its norm upon projection. Therefore, the energy expectation value for projected VQE becomes
\begin{equation}
\label{eq:E}
	{\cal E}({\bm\theta}) =  \frac{\langle \tilde\psi({\bm\theta})| \hat H |\tilde\psi({\bm\theta})\rangle}{\langle \tilde\psi({\bm\theta})|\tilde\psi({\bm\theta})\rangle} = \frac{\int_\Omega d\Omega \; D^{s*}_{mm}(\Omega)  \; \langle  \hat H \hat {\cal U} (\Omega) \rangle_{\bm \theta}}{\int_\Omega d\Omega \; D^{s*}_{mm}(\Omega) \; \langle   \hat {\cal U} (\Omega) \rangle_{\bm \theta}}.
\end{equation}
where $\langle \hat O  \hat {\cal U}(\Omega) \rangle_{\bm \theta} \equiv \langle \psi({\bm \theta})| \hat O \hat {\cal U} (\Omega)  | \psi({\bm \theta})\rangle$.

 It is noteworthy that $|\psi\rangle$ can be any quantum state. If $|\psi\rangle$ is an eigenfunction of $\hat S_z$ with an eigenvalue of $m$ such as UCC, one may {\it a priori} integrate out $\alpha$ and $\gamma$ and use the operator $\hat P$ instead of $\hat {\cal P}$,
\begin{equation}
	\label{eq:simplifiedP}
		\hat { P} \approx \sum_g^{N_g} w_g \hat U_g,
\end{equation}
where we have introduced 
the Gauss-Legendre quadrature using the $N_g$ sampling points
 $\{0\le \beta_g \le \pi: g=1,\cdots,N_g\}$ for rotations $\hat  U_g = e^{-i\beta_g \hat S_y}$ along with accordingly defined weights $w_g$. With this form, one typically requires only a few grid points to achieve $\langle \hat S^2\rangle = s(s+1)$ that is accurate to several decimal points, and the convergence is exponentially fast.\cite{Scuseria11,Lestrange18} 
 To be precise, in this work, we propose to evaluate $\langle \hat O \hat U_g\rangle_{\bm\theta}$ for UCC with a quantum computer and then use a classical adder to evaluate ${\cal E}$, instead of constructing a symmetry-projected ansatz.
 
 %On the one hand, w
 With a classical computer, the cost of evaluating $\langle \hat O\hat U_g\rangle_{\bm\theta}$ scales exponentially if $|\psi({\bm\theta})\rangle$ is an exponential ansatz, such as coupled-cluster,\cite{Qiu17B, Qiu18,Tsuchimochi18,Tsuchimochi19A} since the orbital rotations $\hat U_g$ are still many-body operators that include de-excitations. On the other hand, with a quantum computer, %one can show that 
 it is reduced to a polynomial cost by the Hadamard test with a $\hat U_g$ gate controlled by an ancilla qubit $|q_{\rm anc}\rangle$ (see Appendix \ref{app:MatrixElements}).\cite{Garcia13,Mitarai19, Izmaylov20}
 %Since $\hat U_g$ are unitary, it is convenient to regard $\langle \hat H\hat U_g\rangle_{\bm\theta}$ and $\langle \hat U_g\rangle_{\bm\theta}$ as the Hamiltonian coupling and overlap, respectively, between two quantum states $|\psi({\bm\theta})\rangle$ and $\hat U_g|\psi({\bm\theta})\rangle$. They can be simultaneously measured by entangling the two states through the Hadamard test with a $\hat U_g$ gate controlled by an ancilla qubit $|q_{\rm anc}\rangle$ (see the Supplemental Material).\cite{Garcia13,Mitarai19, Izmaylov20} %At the end of the Hadamard test, one may perform a $Z$-measurement for the ancilla qubit, whose expectation value is easily shown to be ${\rm Re}\langle\hat U_g \rangle_{\bm\theta}$. Furthermore, $\langle \hat H\hat U_g\rangle_{\bm\theta}$ can be evaluated by measuring the Hamiltonian expectation value of the post-measurement state on the system register while appropriately taking into account the result of $|q_{\rm anc}\rangle$,\cite{Parrish19,Huggins20}  (see the Supplemental Material).

\subsubsection{Quantum Circuit}
Now, we consider an efficient circuit that is optimal for the controlled-$U_g$ gate. 
We start by writing $\hat S_y$ in the second quantization as a linear combination of imaginary spin-flip excitations,
\begin{equation}
	\hat S_y	=\frac{1}{2i}\sum_{pq}S_{p\bar q} \hat\tau^{p}_{{\bar q}}.
\end{equation}
where $S_{p\bar q} = \langle \phi_p|\phi_{\bar q}\rangle$ is the overlap between the spatial orbitals of $\alpha$ and $\beta$ spins. It is important to note that although almost all spin-projection methods are based on a broken-symmetry basis of spin-unrestricted HF (UHF), their orbital set is spatially nonorthogonal, i.e. $-1 \le S_{p\bar q} \le 1$.\cite{Schlegel86,Tsuchimochi10B} 
  While the Trotterization can be circumvented for singles,\cite{Kivlichan18} the implementation with UHF can add a non-negligible amount of CNOT gates, which result in an overhead to a quantum circuit. We wish to minimize it.
 
\begin{figure}
\includegraphics[width=0.99\linewidth]{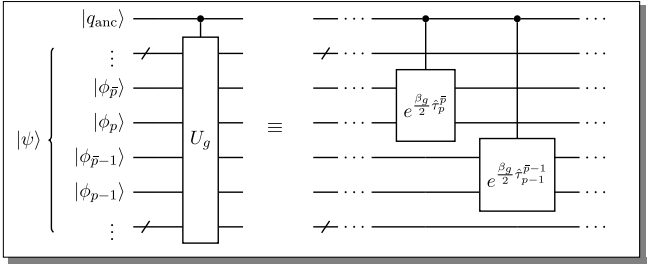}
\caption{The block structure of the controlled-$U_g$ circuit in spin-restricted orbital basis.}\label{fig:CUg}
\end{figure}

\begin{figure*}
\includegraphics[width=0.8\linewidth]{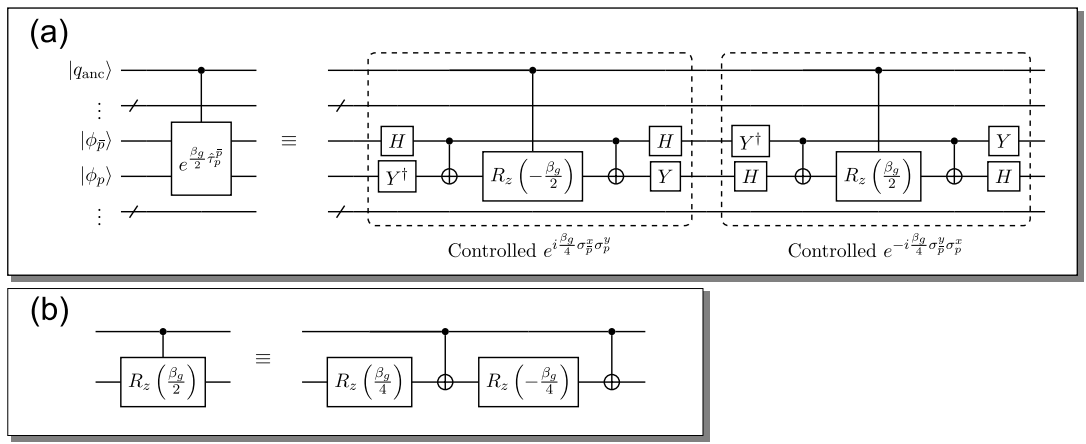}
\caption{Quantum circuits for (a) the spin-flip operation by angle $\beta_g$ and (b) controlled-$R_z$. $H$ and $Y = R_x(-\pi/2)$ respectively transform a qubit to the $\sigma^x$ and $\sigma^y$ basis.  }\label{fig:localSF}
\end{figure*}

In contrast, a quantum circuit for the controlled-$U_g$ gate can be made substantially simpler in an RHF orbital basis, since $S_{p\bar q} \equiv \delta_{pq}$. This gives rise to no Trotter error since each component is commutative with one another. Therefore, 
\begin{equation}
 \label{eq:Ug}
 	\hat U_g = \prod_p^{n/2} e^{\frac{\beta_g}{2} \hat \tau^{\bar p}_{p}}.
 \end{equation}
It can be shown that not only is the number of gates reduced, but also that the controlled-$U_g$ gate with Eq.~(\ref{eq:Ug}) can be efficiently implemented with the Jordan-Wigner transformation.  Suppose spin-orbitals are mapped onto qubits with spins alternating as $\cdots\phi_{p-1}\phi_{\bar p-1}\phi_p\phi_{\bar p}\cdots$. Then, we have, in an RHF orbital basis,
\begin{equation}
\frac{\beta_g}{2} \hat \tau_p^{\bar p} = \frac{i\beta_g}{4}\left(\sigma_{\bar p}^x  \sigma_p^y-\sigma_{\bar p}^y\sigma_p^x \right),
\end{equation} 
which does not entail a sequence of nearest-neighbor CNOT gates that would be needed for a product of $\sigma^z$ between $p$ and ${\bar q}$ to ensure the anti-commutation relation of Fermion operators.\cite{Barkoutsos18, Romero19} Thus, the controlled-$U_g$ gate has a nice block structure with local (controlled) spin-flips $e^{\frac{\beta_g}{2} \hat \tau^{\bar p}_{ p}}$ (Figure \ref{fig:CUg}).
A quantum circuit that performs a local spin-flip is given in Figure \ref{fig:localSF}, along with a controlled-$R_z$ gate decomposed to two $R_z$ and CNOT gates to facilitate actual implementations. With this circuit, the overall overhead introduced by the controlled-$U_g$ gate is negligible, with the number of additional CNOT gates being only 4$n$. Therefore, a large number of measurements required for the purposes of preserving spin-symmetry in both the constrained approach and L\"owdin's spin-projection scheme can be 
completely avoided with our scheme, which needs only $O(N_gn^4)$ measurements. 

 Implementing $e^{-i\alpha\hat S_z}$ is straightforward independent of the basis, thanks to the diagonal nature of $\hat S_z$. A full quantum circuit for $\hat {\cal U} (\Omega)$ is also available in Appendix \ref{app:FullSpin}.

\subsubsection{Broken-Symmetry Ansatz}
While the foregoing discussion holds true with regard to the use of an RHF orbital basis, spin-projection can be even more beneficial if combined with  physically-motivated broken-symmetry ans\"atze.\cite{Tsuchimochi15C,Tsuchimochi16A} One could use BS-UCCSD for this purpose,  but here we describe a general procedure to deliberately break spin-symmetry by applying spin-dependent orbital rotations to a quantum state $|\psi\rangle$ in a way similar to $\hat U_g$. To achieve this task, we conveniently 
 write the orbital rotation operator $\hat K$ defined as a product of spin-dependent Givens rotations $\{e^{\kappa_q^p \hat \tau_q^p}, e^{\kappa_{\bar q}^{\bar p} \hat \tau_{\bar q}^{\bar p}}\}$. 
This procedure has been used  to generate arbitrary Slater determinants. \cite{Wecker15, Kivlichan18} Applying $\hat K$ to RHF generates any UHF state represented by the RHF orbitals. This means that the projected HF (PHF) can be simply given by
\begin{equation}
	|\tilde\psi_{\rm PHF}\rangle = \hat P\hat K|\Phi_{\rm RHF}\rangle.
\end{equation}
It is worth noting that the orbital basis is still that of RHF, whereas UHF is expressed by a linear combination of excited determinants from $|\Phi_{\rm RHF}\rangle$ --- the Thouless theorem.\cite{Thouless60}
We can extend this scheme to any other VQE ansatz including UCC without loss of generality.\cite{Sherrill98,Helgaker00,Mizukami19,comment} For UCCSD, however, $\hat {\cal T}_1$ and $\hat K$ play similar roles and are considered largely redundant. Therefore, in our study we omit $\hat {\cal T}_1$ and consider UCCD combined with $\hat K$ and spin-projection:
\begin{equation}
\label{eq:PUCCD}	|\tilde\psi_{\rm PUCCD}\rangle = \hat P \hat K e^{\hat {\cal T}_2} |\Phi_{\rm RHF}\rangle,
\end{equation}
where both the $\kappa$ and $t$ amplitudes are fully optimized. We call this scheme projected UCCD (PUCCD). Note that the orbital-optimization effect is encoded in $\hat K$. This way, the amplitudes in $\hat {\cal T}_2$ are expected to be small, and higher excitation effects needed for strong correlation are now shifted to the orbital rotation followed by spin-projection. Accordingly, in many cases, $e^{\hat {\cal T}_2}$ in Eq.~(\ref{eq:PUCCD}) is well approximated by $\mu=1$, which is referred to as the disentangled PUCCD (dPUCCD), following the work of Evangelista and co-workers.\cite{Evangelista19} 

HF and UCC are both invariant with respect to orbital rotations within occupied and virtual spaces.\cite{Chen12} Applying spin-projection does not change this property.\cite{Tsuchimochi18} Therefore, it is enough to take into account only the occupied-virtual orbital rotations with $\{\kappa_i^a,\kappa_{\bar i}^{\bar a}\}$. However, introducing the Trotter approximation in $e^{\hat {\cal T}_2}$ of Eq.~(\ref{eq:PUCCD}) no longer guarantees the said invariance. Nevertheless, our experiences indicate that occupied-occupied and virtual-virtual rotations are very much redundant even for dPUCCD, considering that their energy derivatives were found to be numerically zero. Thus, they will not be taken into consideration below. 
\begin{figure}
\includegraphics[width=0.99\linewidth]{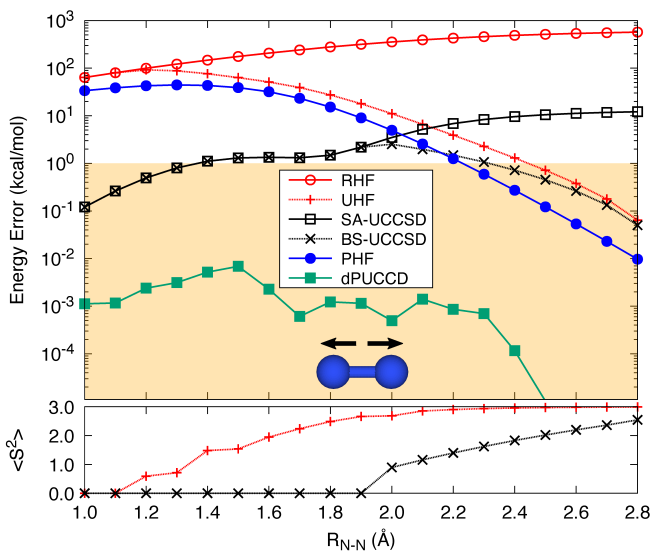}	
\caption{Energy error from FCI in the dissociation of N$_2$. The lower panel shows $\langle \hat S^2 \rangle$ of UHF and BS-UCCSD.}
\label{fig:N2}
\end{figure}

\section{Illustrative calculations}\label{sec:results}
 We have used PySCF\cite{pyscf} to generate molecular integrals, OpenFermion\cite{openfermion} to perform Jordan-Wigner transformation of $\hat H$ and $\hat S^2$, and Qulacs\cite{Qulacs} to construct quantum circuits for simulation on a classical computer. For optimization, the L-BFGS method was used.\cite{scipy} 
$N_g$ required for the numerically exact projection varies depending on the degree of spin-contamination; we have used $N_g = 2$ for N$_2$ and H$_2$O and $N_g = 3$ for O, which are found to be sufficient for the $\langle \hat S^2 \rangle$ values to be accurate to at least $10^{-10}$. The exponential operators in Trotterization  are applied in the following order: $abij \rightarrow \bar ab\bar ij \rightarrow\bar a\bar b\bar i\bar j \rightarrow ai \rightarrow \bar a \bar i$, where, in each spin-block, the leftmost virtual label is the outermost loop while the rightmost occupied label is the innermost loop. The Trotter approximation is considered as having been converged with $\mu=10$.

We first assess the improvement that spin-projection has to offer by computing the potential energy curves of the triple bond dissociation of  the singlet N$_2$ ($m=0$).   We have used the STO-6G basis set and frozen $1s$ and $2s$ core electrons of atoms, resulting in an active space of $(6e,6o)$. 
Plotted in Figure \ref{fig:N2} are the energy errors from FCI (kcal/mol) for  N$_2$ in a logarithmic scale. The shaded area corresponds to the ``chemical accuracy,’’ i.e., errors less than 1 kcal/mol.
  In the lower panel, $\langle \hat S^2\rangle$ of UHF and BS-UCCSD are likewise depicted. All other methods numerically preserve the correct singlet value $\langle \hat S^2 \rangle = 0$ and are, therefore, not shown. The energy of BS-UCCSD starts deviating from that of SA-UCCSD by breaking the symmetry restriction in amplitudes.
It is found that this so-called Coulson-Fischer point happens at a much longer distance for UCCSD (2.0 {\AA}) compared to that of HF (1.2 {\AA}). This is because strong correlation can be partially captured via exponential excitations in UCCSD, thereby mitigating spin contamination. However, the ansatz only accounts for disconnected higher excitations, e.g., $\hat {\cal T}_2^2$, resulting in larger errors. On the other hand, it is worth noting that dPUCCD delivers essentially exact results for these systems while guaranteeing the singlet spin; the largest deviation is 0.007 kcal/mol at 1.5 \AA. The error is at least one order of magnitude smaller than those of SA-UCCSD and BS-UCCSD.

\begin{figure}[t]
\includegraphics[width=0.98\linewidth]{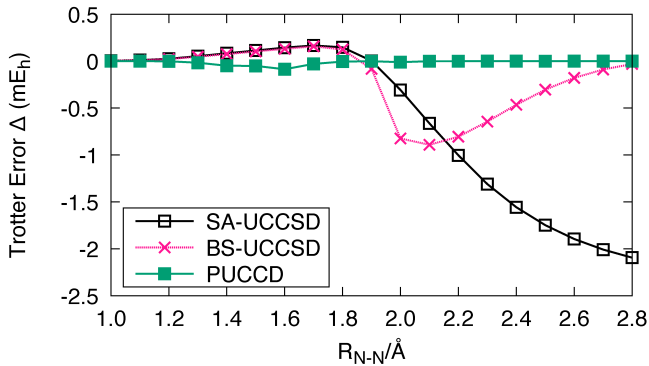}
\caption{Trotter error in energy  $\Delta$ of disentangled UCC from $\mu=10$ results for N$_2$.}\label{fig:Trotter_error}
\end{figure}

\begin{table}[b]
\caption{Norms of singles and doubles amplitudes for N$_2$ at 2.2 \AA. The numbers in parentheses indicate the distance from the converged amplitudes of disentangled UCC ($\mu=1$).}\label{tab:amp}
\begin{tabular}{cccc}
\hline\hline
&  SA-UCCSD & BS-UCCSD   & PUCCD\\
\hline
Singles & $9\times 10^{-6}$ ($1\times 10^{-5}$) & 1.50 (0.14) & 1.75 ($3\times 10^{-5}$)\\
Doubles & 1.13 (0.072) &  0.44 (0.11) &  0.10 ($6\times 10^{-5}$) \\
\hline\hline
\end{tabular}
    
\end{table}

\begin{table}[t]
\caption{Energy errors from FCI (kcal/mol) for disentangled methods using the mixed basis (6-31G and STO-6G for the oxygen and hydrogen atoms). Numbers in the parentheses are the errors obtained with the minimal STO-6G basis.}\label{tb:H2O}
\begin{tabular}{cccccc}
\hline\hline
R$_{\rm O-H}$ (\AA)&   SA-UCCSD & BS-UCCSD	 & PUCCD \\
\hline\hline
1.0 & 	0.4 (0.0)&	0.4 (0.0) &	 0.1 (0.0)\\
1.5 &	2.2	(0.2)&	2.2 (0.1)&0.3 (0.0)\\
2.0 & 	4.9	(0.7) & 4.9 (0.7)& 0.3 (0.0) \\
2.5 &	8.4 (3.4)&	1.9 (0.4)&	0.1 (0.0)  \\
\hline\hline
\end{tabular}
\end{table}

Both in PHF and PUCCD, higher excitations are treated via spin-projection and orbital-rotations $\hat K$, which do not need Trotterization. Hence, it is expected that the doubles amplitudes of PUCCD are small, thereby allowing for a small Trotter error in dPUCCD compared to those of the disentangled UCCSD (dUCCSD). This is demonstrated in Figure \ref{fig:Trotter_error}, where we have plotted the Trotter error $\Delta$ of each disentangled scheme in the dissociation of N$_2$. In Table \ref{tab:amp}, we tabulated the norms of singles (either $|{\bf t}|$ or $|{\bm\kappa}|$) and  doubles amplitudes obtained at $R_{\rm N-N} = 2.2$ {\AA} with $\mu=10$ as well as the distances from those of $\mu=1$, as another indicator of the Trotter error. As can be clearly seen, PUCCD gives the minimal $T_2$ amplitudes, thereby allowing for an efficient treatment of strong correlation. In contrast, the converged $T_1$-amplitudes of BS-UCCSD are considerable and require a re-optimization of amplitudes for different $\mu$. This indicates that BS-dUCCSD is a whole different ansatz from BS-UCCSD, even for a large bond length, $R_{\rm N-N} =2.8 $ {\AA}, where both yield almost identical energies, as is evident from the totally different $\langle \hat S^2\rangle$, i.e., 1.37 for $\mu=1$ and 2.55 for $\mu=10$.  

  The above test system employs the minimal basis and admittedly neglects most of the dynamical correlation effect. To show that the accuracy of PUCCD is not a fortuitous artifact resulting from the chosen basis, we have also carried out the calculations for the symmetric bond dissociation of H$_2$O with a fixed angle of 104.5$^\circ$ using  6-31G and STO-6G for oxygen and hydrogen, respectively. 
Table \ref{tb:H2O} summarizes the energy errors of each disentangled method from FCI, where the numbers in parentheses imply those obtained with the STO-6G basis for all the atoms for comparison. The presence of non-negligible dynamical correlation leads to substantial errors if strong correlation is not treated appropriately. This is the case for SA- and BS-UCCSD, while PUCCD still remains to  achieve the chemical accuracy.

Lastly, we computed the singlet-triplet energy gap of the oxygen atom to show that neglecting spin-symmetry can often result in a catastrophic error. The basis set used is 6-31G with $1s$ frozen core, constituting another example with a balanced mixture of dynamical and strong correlation effects. Table \ref{tab:o} lists the singlet and triplet energies in Hartree, computed with $m=0$ and $2$, respectively, along with their difference $\Delta_{\rm ST}$ in kcal/mol. It is readily clear that both SA- and BS-UCCSD schemes failed to describe strong correlations of the open-shell singlet oxygen. BS-UCCSD with $m=0$ spontaneously breaks spin and ends up in mainly the low-spin triplet state ($\langle \hat S^2\rangle =1.96$). Spin-projection in dPUCCD enables to capture strong correlation in the singlet oxygen.  As a result, its $\Delta_{\rm ST}$ is found to be in excellent agreement with FCI.
\begin{table}[b]
\caption{Singlet and triplet energies (Hartree) and the spin gap (kcal/mol) of the oxygen atom.}\label{tab:o}
\begin{tabular}{ccccc}
\hline\hline
& FCI &   SA-UCCSD & BS-UCCSD	 & dPUCCD\\
\hline
Singlet &-74.756 28 &	-74.748 53 &	-74.826 40 &	-74.756 07\\
Triplet &-74.838 56 &	-74.838 14 &	-74.838 14 &	-74.838 17\\
$\Delta_{\rm ST}$&51.6&	56.2&	7.37 &	51.5\\
\hline\hline
\end{tabular}
\end{table}

\section{Conclusions}\label{sec:conclustions}
In this Letter, we have shown that numerically exact spin-projection can be made feasible within the framework of VQE. It should be emphasized that spin-projection attains remarkable accuracy in exchange for one ancilla qubit, albeit with little overhead in circuit depth (additional $4n$ CNOT gates). We note that one disadvantage of our scheme is the increase in the number of measurements, but it grows only linearly with $N_g$ as opposed to standard nonorthogonal methods that show a quadratic scaling.\cite{Huggins20} Also, the scheme can potentially suffer from errors when a projected state is not suitable for dealing with strong correlation in the system \cite{Qiu17A, Degroote16}. Nonetheless, we should once again stress that the spin-projection operator lends itself straightforwardly to {\it any ansatz} in order to guarantee the desired spin-symmetry, and therefore is a promising, versatile tool.

{\it Acknowledgements}---This work was supported by JSPS KAKENHI grant numbers JP18H03900 and JP20K15231.

\appendix

\section{Matrix elements}\label{app:MatrixElements}
\begin{figure}[t]
	\includegraphics[width=0.8\linewidth]{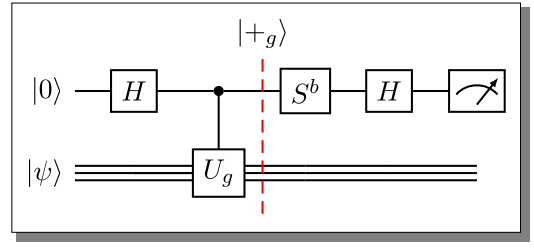}
	\caption{Hadamard test to obtain the real ($b=0$) and imaginary ($b=1$) parts of expectation value $\langle \psi|\hat U_g |\psi\rangle$.}
	\label{fig:Hadamard}
\end{figure}

It is now well known that the expectation value of a unitary operator can be evaluated by the Hadamard test. Here, we review the basic idea and derive the Hamiltonian and overlap matrix elements between $|\psi\rangle$ and $\hat U_g |\psi\rangle$. Figure \ref{fig:Hadamard} presents a quantum circuit to generate an entangled state
\begin{align}
	|+_g\rangle = \frac{1}{\sqrt{2}}\left(|0\rangle \otimes |\psi\rangle  + |1\rangle\otimes \hat U_g |\psi\rangle\right)
\end{align}
where the first qubit is an ancilla qubit $|q_{\rm anc}\rangle$ initially set to $|0\rangle$, and the rest $n$ qubits constitutes the system register to represent a broken-symmetry wave function ansatz $|\psi\rangle$. Applying a Hadamard gate to the ancilla qubit further gives 
\begin{align}
 H_{\rm anc} |+_g\rangle = \frac{1}{2}\left[|0\rangle \otimes \left(|\psi\rangle + \hat U_g |\psi\rangle\right) +|1\rangle \otimes \left(|\psi\rangle - \hat U_g |\psi\rangle\right)\right].
\end{align}
Now one may perform a $Z$-measurement for the ancilla qubit, whose expectation value is easily shown to be ${\rm Re}\langle \psi|\hat U_g |\psi\rangle$: the probabilities of the ancilla qubit found to be $|0\rangle$ and $|1\rangle$ are, respectively, $p_0 =\left\|\frac{1}{2}\left(|\psi\rangle + \hat U_g |\psi\rangle\right)\right\|^2 = \frac{1}{2} \left(1+{\rm Re}\langle \psi|\hat U_g |\psi\rangle\right)$ and similarly $p_1= \frac{1}{2} \left(1-{\rm Re}\langle \psi|\hat U_g |\psi\rangle\right)$, and therefore
\begin{align}
	\langle \hat \sigma^z_{\rm anc}\rangle &=(+1) p_0 + (-1) p_1 \nonumber\\
%	&= \frac{1}{2} \left(1+{\rm Re}\langle \psi|\hat U_g |\psi\rangle\right) - \frac{1}{2} \left(1-{\rm Re}\langle \psi|\hat U_g |\psi\rangle\right) \nonumber\\
	&
	= {\rm Re}\langle \psi|\hat U_g |\psi\rangle.\label{eq:Ug_expectation}
\end{align}
The post-measurement state on the system register $|\psi_{\rm post}\rangle$ becomes, depending on the result of the ancilla qubit measurement,
\begin{align}
	\begin{cases}
	|\psi_{\rm post}^{(0)}\rangle =\displaystyle \frac{|\psi\rangle + \hat U_g |\psi\rangle}{\sqrt{2\left(1+{\rm Re}\langle \psi|\hat U_g |\psi\rangle\right)}}
		& (|q_{\rm anc}\rangle = |0\rangle)\\
|\psi_{\rm post}^{(1)}\rangle =\displaystyle \frac{|\psi\rangle - \hat U_g |\psi\rangle}{\sqrt{2\left(1-{\rm Re}\langle \psi|\hat U_g |\psi\rangle\right)}}
		& (|q_{\rm anc}\rangle = |1\rangle)
	\end{cases}
\end{align}
If one simply measures the expectation value of $\hat H$ with the post-measurement state, one finds
\begin{align}
	\langle \psi_{\rm post}^{(0)}| \hat H |\psi_{\rm post}^{(0)}\rangle = 		\displaystyle \frac{\langle\psi|\hat H|\psi\rangle + \langle\psi|\hat U_g^\dag \hat H\hat U_g|\psi\rangle + 2{\rm Re} \langle \psi|\hat H\hat U_g |\psi\rangle}{2\left(1+{\rm Re}\langle \psi|\hat U_g |\psi\rangle\right)}
\\
	\langle \psi_{\rm post}^{(1)}| \hat H |\psi_{\rm post}^{(1)}\rangle = 
		\displaystyle \frac{\langle\psi|\hat H|\psi\rangle + \langle\psi|\hat U_g^\dag \hat H\hat U_g|\psi\rangle - 2{\rm Re} \langle \psi|\hat H\hat U_g |\psi\rangle}{2\left(1-{\rm Re}\langle \psi|\hat U_g |\psi\rangle\right)}	
		\end{align}
Hence, $\langle \hat H\hat U_g\rangle$ can be neatly evaluated by appropriately taking into account the measurement of $\hat \sigma^z_{\rm anc}$ ($+1$ with probability $p_0$ or $-1$ with probability $p_1$),
\begin{align}
	\langle  \hat \sigma^z_{\rm anc} \otimes \hat H \rangle
	= &(+1) p_0  \langle \psi_{\rm post}^{(0)}| \hat H |\psi_{\rm post}^{(0)}\rangle\nonumber\\
	 &+ (-1) p_1	\langle \psi_{\rm post}^{(1)}| \hat H |\psi_{\rm post}^{(1)}\rangle \nonumber\\
&= {\rm Re} \langle \psi|\hat H\hat U_g |\psi\rangle \label{eq:HUg_expectaion}
\end{align}

The above scheme only permits for the evaluation of the real parts of expectation values. The imaginary parts ${\rm Im}\langle \psi|\hat H \hat U_g|\psi\rangle$ and ${\rm Im}\langle \psi|\hat U_g|\psi\rangle$ can be obtained by acting the phase operator $S=\begin{pmatrix}
	1 & 0 \\ 0 & i
\end{pmatrix}$ to the ancilla qubit after the controlled-$U_g$ gate in the Hadamard test (see Figure \ref{fig:Hadamard}). Having said that, for the simplified spin-projection scheme as given in the main text, all the quantities come out real as long as the molecular orbitals are real.

\begin{figure*}[t]
\includegraphics[width=0.8\linewidth]{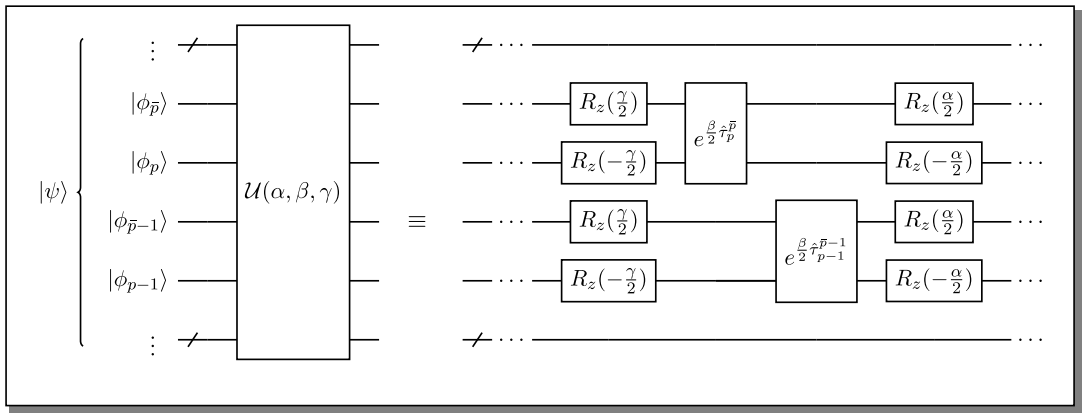}
\caption{Quantum circuit for $\hat {\cal U}(\alpha,\beta,\gamma)$.}\label{fig:U}
\end{figure*}
\section{Quantum circuit for the full-spin rotation}\label{app:FullSpin}
For the full-spin projection for a spin-generalized quantum state (both $\hat S^2$ and $\hat S_z$ symmetries are broken), one has to implement $\hat {\cal U}(\Omega)$ (Eq.~(\ref{eq:UOmega})) instead of the simplified rotator $\hat U_g$. This necessitates a quantum circuit that performs $e^{-i\theta\hat S_z}$ for $\theta = \alpha,\gamma$, where
\begin{align}
	\hat S_z = \frac{1}{2}\sum_p \left(a_p^\dag a_p - a_{\bar p}^\dag a_{\bar p}\right)
\end{align}
The diagonal nature of $\hat S_z$ makes it particularly easy to implement its exponential form on a quantum computer. Namely,
\begin{align}
	e^{-i\theta\hat S_z} &= e^{-i\frac{\theta}{2}\sum_p \left(a_p^\dag a_p - a_{\bar p}^\dag a_{\bar p}\right)	}\nonumber\\&
	\equiv \prod_p e^{-i\frac{\theta}{2} a_p^\dag a_p} e^{i\frac{\theta}{2} a_{\bar p}^\dag a_{\bar p}	} \nonumber\\
%	&=\left(\prod_p e^{-i\frac{\theta}{4} (a_p^\dag a_p - a_p a_p^\dag) }\right) \left( \prod_p e^{i\frac{\theta}{4} (a_{\bar p}^\dag a_{\bar p}	-a_{\bar p}a_{\bar p}^\dag)}\right) \nonumber\\
	&= \prod_p e^{-i\frac{\theta}{4} (I_p - \sigma_p^z)} 
	    e^{i\frac{\theta}{4} (I_{\bar p} - \sigma_{\bar p}^z)} \nonumber\\
	&= \prod_p e^{i\frac{\theta}{4} \sigma_p^z}
	 e^{-i\frac{\theta}{4}  \sigma_{\bar p}^z} 
\end{align}
This means that $e^{-i\theta\hat S_z}$ only requires single qubit operations, where all qubits that represent up-spin and down-spin orbitals are rotated by $R_z\left( -\frac{\theta}{2} \right)$ and $R_z\left( \frac{\theta}{2} \right)$ , respectively. Therefore, a quantum circuit that perform the full spin-projection with $\hat {\cal U}(\alpha,\beta,\gamma)$ simply becomes as follows:
\begin{enumerate}
	\item Perform single qubit operations $R_z\left(- \frac{\gamma}{2} \right)$ and $R_z\left(\frac{\gamma}{2} \right)$ for up-spin and down-spin qubits.
	\item Perform local spin-flips $e^{i\frac{\beta}{4}(\sigma_p^y \sigma_{\bar p}^x - \sigma_p^x \sigma_{\bar p}^y) }$
	\item Again, perform single qubit operations $R_z\left(-\frac{\alpha}{2} \right)$ and $R_z\left(\frac{\alpha}{2} \right)$ for up-spin and down-spin qubits. 
\end{enumerate}
We show our quantum circuit for $\hat {\cal U}(\alpha,\beta,\gamma)$ in Figure \ref{fig:U}.


\begin{thebibliography}{53}%
\makeatletter
\providecommand \@ifxundefined [1]{%
 \@ifx{#1\undefined}
}%
\providecommand \@ifnum [1]{%
 \ifnum #1\expandafter \@firstoftwo
 \else \expandafter \@secondoftwo
 \fi
}%
\providecommand \@ifx [1]{%
 \ifx #1\expandafter \@firstoftwo
 \else \expandafter \@secondoftwo
 \fi
}%
\providecommand \natexlab [1]{#1}%
\providecommand \enquote  [1]{``#1''}%
\providecommand \bibnamefont  [1]{#1}%
\providecommand \bibfnamefont [1]{#1}%
\providecommand \citenamefont [1]{#1}%
\providecommand \href@noop [0]{\@secondoftwo}%
\providecommand \href [0]{\begingroup \@sanitize@url \@href}%
\providecommand \@href[1]{\@@startlink{#1}\@@href}%
\providecommand \@@href[1]{\endgroup#1\@@endlink}%
\providecommand \@sanitize@url [0]{\catcode `\\12\catcode `\$12\catcode
  `\&12\catcode `\#12\catcode `\^12\catcode `\_12\catcode `\%12\relax}%
\providecommand \@@startlink[1]{}%
\providecommand \@@endlink[0]{}%
\providecommand \url  [0]{\begingroup\@sanitize@url \@url }%
\providecommand \@url [1]{\endgroup\@href {#1}{\urlprefix }}%
\providecommand \urlprefix  [0]{URL }%
\providecommand \Eprint [0]{\href }%
\providecommand \doibase [0]{http://dx.doi.org/}%
\providecommand \selectlanguage [0]{\@gobble}%
\providecommand \bibinfo  [0]{\@secondoftwo}%
\providecommand \bibfield  [0]{\@secondoftwo}%
\providecommand \translation [1]{[#1]}%
\providecommand \BibitemOpen [0]{}%
\providecommand \bibitemStop [0]{}%
\providecommand \bibitemNoStop [0]{.\EOS\space}%
\providecommand \EOS [0]{\spacefactor3000\relax}%
\providecommand \BibitemShut  [1]{\csname bibitem#1\endcsname}%
\let\auto@bib@innerbib\@empty
%</preamble>
\bibitem [{\citenamefont {Peruzzo}\ \emph {et~al.}(2014)\citenamefont
  {Peruzzo}, \citenamefont {McClean}, \citenamefont {Shadbolt}, \citenamefont
  {Yung}, \citenamefont {Zhou}, \citenamefont {Love}, \citenamefont
  {Aspuru-Guzik},\ and\ \citenamefont {O'Brien}}]{Peruzzo14}%
  \BibitemOpen
  \bibfield  {author} {\bibinfo {author} {\bibfnamefont {A.}~\bibnamefont
  {Peruzzo}}, \bibinfo {author} {\bibfnamefont {J.}~\bibnamefont {McClean}},
  \bibinfo {author} {\bibfnamefont {P.}~\bibnamefont {Shadbolt}}, \bibinfo
  {author} {\bibfnamefont {M.-H.}\ \bibnamefont {Yung}}, \bibinfo {author}
  {\bibfnamefont {X.-Q.}\ \bibnamefont {Zhou}}, \bibinfo {author}
  {\bibfnamefont {P.~J.}\ \bibnamefont {Love}}, \bibinfo {author}
  {\bibfnamefont {A.}~\bibnamefont {Aspuru-Guzik}}, \ and\ \bibinfo {author}
  {\bibfnamefont {J.~L.}\ \bibnamefont {O'Brien}},\ }\href@noop {} {\bibfield
  {journal} {\bibinfo  {journal} {Nat. Comm.}\ }\textbf {\bibinfo {volume}
  {5}},\ \bibinfo {pages} {4213} (\bibinfo {year} {2014})}\BibitemShut
  {NoStop}%
\bibitem [{\citenamefont {O'Malley}\ \emph {et~al.}(2016)\citenamefont
  {O'Malley}, \citenamefont {Babbush}, \citenamefont {Kivlichan}, \citenamefont
  {Romero}, \citenamefont {McClean}, \citenamefont {Barends}, \citenamefont
  {Kelly}, \citenamefont {Roushan}, \citenamefont {Tranter}, \citenamefont
  {Ding}, \citenamefont {Campbell}, \citenamefont {Chen}, \citenamefont {Chen},
  \citenamefont {Chiaro}, \citenamefont {Dunsworth}, \citenamefont {Fowler},
  \citenamefont {Jeffrey}, \citenamefont {Lucero}, \citenamefont {Megrant},
  \citenamefont {Mutus}, \citenamefont {Neeley}, \citenamefont {Neill},
  \citenamefont {Quintana}, \citenamefont {Sank}, \citenamefont {Vainsencher},
  \citenamefont {Wenner}, \citenamefont {White}, \citenamefont {Coveney},
  \citenamefont {Love}, \citenamefont {Neven}, \citenamefont {Aspuru-Guzik},\
  and\ \citenamefont {Martinis}}]{OMalley16}%
  \BibitemOpen
  \bibfield  {author} {\bibinfo {author} {\bibfnamefont {P.~J.~J.}\
  \bibnamefont {O'Malley}}, \bibinfo {author} {\bibfnamefont {R.}~\bibnamefont
  {Babbush}}, \bibinfo {author} {\bibfnamefont {I.~D.}\ \bibnamefont
  {Kivlichan}}, \bibinfo {author} {\bibfnamefont {J.}~\bibnamefont {Romero}},
  \bibinfo {author} {\bibfnamefont {J.~R.}\ \bibnamefont {McClean}}, \bibinfo
  {author} {\bibfnamefont {R.}~\bibnamefont {Barends}}, \bibinfo {author}
  {\bibfnamefont {J.}~\bibnamefont {Kelly}}, \bibinfo {author} {\bibfnamefont
  {P.}~\bibnamefont {Roushan}}, \bibinfo {author} {\bibfnamefont
  {A.}~\bibnamefont {Tranter}}, \bibinfo {author} {\bibfnamefont
  {N.}~\bibnamefont {Ding}}, \bibinfo {author} {\bibfnamefont {B.}~\bibnamefont
  {Campbell}}, \bibinfo {author} {\bibfnamefont {Y.}~\bibnamefont {Chen}},
  \bibinfo {author} {\bibfnamefont {Z.}~\bibnamefont {Chen}}, \bibinfo {author}
  {\bibfnamefont {B.}~\bibnamefont {Chiaro}}, \bibinfo {author} {\bibfnamefont
  {A.}~\bibnamefont {Dunsworth}}, \bibinfo {author} {\bibfnamefont {A.~G.}\
  \bibnamefont {Fowler}}, \bibinfo {author} {\bibfnamefont {E.}~\bibnamefont
  {Jeffrey}}, \bibinfo {author} {\bibfnamefont {E.}~\bibnamefont {Lucero}},
  \bibinfo {author} {\bibfnamefont {A.}~\bibnamefont {Megrant}}, \bibinfo
  {author} {\bibfnamefont {J.~Y.}\ \bibnamefont {Mutus}}, \bibinfo {author}
  {\bibfnamefont {M.}~\bibnamefont {Neeley}}, \bibinfo {author} {\bibfnamefont
  {C.}~\bibnamefont {Neill}}, \bibinfo {author} {\bibfnamefont
  {C.}~\bibnamefont {Quintana}}, \bibinfo {author} {\bibfnamefont
  {D.}~\bibnamefont {Sank}}, \bibinfo {author} {\bibfnamefont {A.}~\bibnamefont
  {Vainsencher}}, \bibinfo {author} {\bibfnamefont {J.}~\bibnamefont {Wenner}},
  \bibinfo {author} {\bibfnamefont {T.~C.}\ \bibnamefont {White}}, \bibinfo
  {author} {\bibfnamefont {P.~V.}\ \bibnamefont {Coveney}}, \bibinfo {author}
  {\bibfnamefont {P.~J.}\ \bibnamefont {Love}}, \bibinfo {author}
  {\bibfnamefont {H.}~\bibnamefont {Neven}}, \bibinfo {author} {\bibfnamefont
  {A.}~\bibnamefont {Aspuru-Guzik}}, \ and\ \bibinfo {author} {\bibfnamefont
  {J.~M.}\ \bibnamefont {Martinis}},\ }\href {\doibase
  10.1103/PhysRevX.6.031007} {\bibfield  {journal} {\bibinfo  {journal} {Phys.
  Rev. X}\ }\textbf {\bibinfo {volume} {6}},\ \bibinfo {pages} {031007}
  (\bibinfo {year} {2016})}\BibitemShut {NoStop}%
\bibitem [{\citenamefont {Kutzelnigg}(1977)}]{Kutzelnigg77}%
  \BibitemOpen
  \bibfield  {author} {\bibinfo {author} {\bibfnamefont {W.}~\bibnamefont
  {Kutzelnigg}},\ }\enquote {\bibinfo {title} {Pair correlation theories},}\
  in\ \href@noop {} {\emph {\bibinfo {booktitle} {Methods of Electronic
  Structure Theory}}},\ \bibinfo {editor} {edited by\ \bibinfo {editor}
  {\bibfnamefont {H.~F.}\ \bibnamefont {Schaefer}}}\ (\bibinfo  {publisher}
  {Springer US},\ \bibinfo {address} {Boston, MA},\ \bibinfo {year} {1977})\
  pp.\ \bibinfo {pages} {129--188}\BibitemShut {NoStop}%
\bibitem [{\citenamefont {Bartlett}\ \emph {et~al.}(1989)\citenamefont
  {Bartlett}, \citenamefont {Kucharski},\ and\ \citenamefont
  {Noga}}]{Bartlett89}%
  \BibitemOpen
  \bibfield  {author} {\bibinfo {author} {\bibfnamefont {R.~J.}\ \bibnamefont
  {Bartlett}}, \bibinfo {author} {\bibfnamefont {S.~A.}\ \bibnamefont
  {Kucharski}}, \ and\ \bibinfo {author} {\bibfnamefont {J.}~\bibnamefont
  {Noga}},\ }\href@noop {} {\bibfield  {journal} {\bibinfo  {journal} {Chem.
  Phys. Lett.}\ }\textbf {\bibinfo {volume} {155}},\ \bibinfo {pages} {133 }
  (\bibinfo {year} {1989})}\BibitemShut {NoStop}%
\bibitem [{\citenamefont {Taube}\ and\ \citenamefont
  {Bartlett}(2006)}]{Taube06}%
  \BibitemOpen
  \bibfield  {author} {\bibinfo {author} {\bibfnamefont {A.~G.}\ \bibnamefont
  {Taube}}\ and\ \bibinfo {author} {\bibfnamefont {R.~J.}\ \bibnamefont
  {Bartlett}},\ }\href {\doibase 10.1002/qua.21198} {\bibfield  {journal}
  {\bibinfo  {journal} {Int. J. Quantum Chem.}\ }\textbf {\bibinfo {volume}
  {106}},\ \bibinfo {pages} {3393} (\bibinfo {year} {2006})}\BibitemShut
  {NoStop}%
\bibitem [{\citenamefont {Barkoutsos}\ \emph {et~al.}(2018)\citenamefont
  {Barkoutsos}, \citenamefont {Gonthier}, \citenamefont {Sokolov},
  \citenamefont {Moll}, \citenamefont {Salis}, \citenamefont {Fuhrer},
  \citenamefont {Ganzhorn}, \citenamefont {Egger}, \citenamefont {Troyer},
  \citenamefont {Mezzacapo}, \citenamefont {Filipp},\ and\ \citenamefont
  {Tavernelli}}]{Barkoutsos18}%
  \BibitemOpen
  \bibfield  {author} {\bibinfo {author} {\bibfnamefont {P.~K.}\ \bibnamefont
  {Barkoutsos}}, \bibinfo {author} {\bibfnamefont {J.~F.}\ \bibnamefont
  {Gonthier}}, \bibinfo {author} {\bibfnamefont {I.}~\bibnamefont {Sokolov}},
  \bibinfo {author} {\bibfnamefont {N.}~\bibnamefont {Moll}}, \bibinfo {author}
  {\bibfnamefont {G.}~\bibnamefont {Salis}}, \bibinfo {author} {\bibfnamefont
  {A.}~\bibnamefont {Fuhrer}}, \bibinfo {author} {\bibfnamefont
  {M.}~\bibnamefont {Ganzhorn}}, \bibinfo {author} {\bibfnamefont {D.~J.}\
  \bibnamefont {Egger}}, \bibinfo {author} {\bibfnamefont {M.}~\bibnamefont
  {Troyer}}, \bibinfo {author} {\bibfnamefont {A.}~\bibnamefont {Mezzacapo}},
  \bibinfo {author} {\bibfnamefont {S.}~\bibnamefont {Filipp}}, \ and\ \bibinfo
  {author} {\bibfnamefont {I.}~\bibnamefont {Tavernelli}},\ }\href@noop {}
  {\bibfield  {journal} {\bibinfo  {journal} {Phys. Rev. A}\ }\textbf {\bibinfo
  {volume} {98}},\ \bibinfo {pages} {022322} (\bibinfo {year}
  {2018})}\BibitemShut {NoStop}%
\bibitem [{\citenamefont {Moll}\ \emph {et~al.}(2018)\citenamefont {Moll},
  \citenamefont {Barkoutsos}, \citenamefont {Bishop}, \citenamefont {Chow},
  \citenamefont {Cross}, \citenamefont {Egger}, \citenamefont {Filipp},
  \citenamefont {Fuhrer}, \citenamefont {Gambetta}, \citenamefont {Ganzhorn},
  \citenamefont {Kandala}, \citenamefont {Mezzacapo}, \citenamefont {M\"uller},
  \citenamefont {Riess}, \citenamefont {Salis}, \citenamefont {Smolin},
  \citenamefont {Tavernelli},\ and\ \citenamefont {Temme}}]{Moll18}%
  \BibitemOpen
  \bibfield  {author} {\bibinfo {author} {\bibfnamefont {N.}~\bibnamefont
  {Moll}}, \bibinfo {author} {\bibfnamefont {P.}~\bibnamefont {Barkoutsos}},
  \bibinfo {author} {\bibfnamefont {L.~S.}\ \bibnamefont {Bishop}}, \bibinfo
  {author} {\bibfnamefont {J.~M.}\ \bibnamefont {Chow}}, \bibinfo {author}
  {\bibfnamefont {A.}~\bibnamefont {Cross}}, \bibinfo {author} {\bibfnamefont
  {D.~J.}\ \bibnamefont {Egger}}, \bibinfo {author} {\bibfnamefont
  {S.}~\bibnamefont {Filipp}}, \bibinfo {author} {\bibfnamefont
  {A.}~\bibnamefont {Fuhrer}}, \bibinfo {author} {\bibfnamefont {J.~M.}\
  \bibnamefont {Gambetta}}, \bibinfo {author} {\bibfnamefont {M.}~\bibnamefont
  {Ganzhorn}}, \bibinfo {author} {\bibfnamefont {A.}~\bibnamefont {Kandala}},
  \bibinfo {author} {\bibfnamefont {A.}~\bibnamefont {Mezzacapo}}, \bibinfo
  {author} {\bibfnamefont {P.}~\bibnamefont {M\"uller}}, \bibinfo {author}
  {\bibfnamefont {W.}~\bibnamefont {Riess}}, \bibinfo {author} {\bibfnamefont
  {G.}~\bibnamefont {Salis}}, \bibinfo {author} {\bibfnamefont
  {J.}~\bibnamefont {Smolin}}, \bibinfo {author} {\bibfnamefont
  {I.}~\bibnamefont {Tavernelli}}, \ and\ \bibinfo {author} {\bibfnamefont
  {K.}~\bibnamefont {Temme}},\ }\href {\doibase 10.1088/2058-9565/aab822}
  {\bibfield  {journal} {\bibinfo  {journal} {Quantum Sci. Technol.}\ }\textbf
  {\bibinfo {volume} {3}},\ \bibinfo {pages} {030503} (\bibinfo {year}
  {2018})}\BibitemShut {NoStop}%
\bibitem [{\citenamefont {Romero}\ \emph {et~al.}(2019)\citenamefont {Romero},
  \citenamefont {Babbush}, \citenamefont {McClean}, \citenamefont {Hempe},
  \citenamefont {Love},\ and\ \citenamefont {Aspuru-Guzik}}]{Romero19}%
  \BibitemOpen
  \bibfield  {author} {\bibinfo {author} {\bibfnamefont {J.}~\bibnamefont
  {Romero}}, \bibinfo {author} {\bibfnamefont {R.}~\bibnamefont {Babbush}},
  \bibinfo {author} {\bibfnamefont {J.~R.}\ \bibnamefont {McClean}}, \bibinfo
  {author} {\bibfnamefont {C.}~\bibnamefont {Hempe}}, \bibinfo {author}
  {\bibfnamefont {P.~J.}\ \bibnamefont {Love}}, \ and\ \bibinfo {author}
  {\bibfnamefont {A.}~\bibnamefont {Aspuru-Guzik}},\ }\href@noop {} {\bibfield
  {journal} {\bibinfo  {journal} {Quantum Sci. Technol.}\ }\textbf {\bibinfo
  {volume} {4}},\ \bibinfo {pages} {014008} (\bibinfo {year}
  {2019})}\BibitemShut {NoStop}%
\bibitem [{\citenamefont {Evangelista}\ \emph {et~al.}(2019)\citenamefont
  {Evangelista}, \citenamefont {Chan},\ and\ \citenamefont
  {Scuseria}}]{Evangelista19}%
  \BibitemOpen
  \bibfield  {author} {\bibinfo {author} {\bibfnamefont {F.~A.}\ \bibnamefont
  {Evangelista}}, \bibinfo {author} {\bibfnamefont {G.~K.-L.}\ \bibnamefont
  {Chan}}, \ and\ \bibinfo {author} {\bibfnamefont {G.~E.}\ \bibnamefont
  {Scuseria}},\ }\href {\doibase 10.1063/1.5133059} {\bibfield  {journal}
  {\bibinfo  {journal} {J. Chem. Phys.}\ }\textbf {\bibinfo {volume} {151}},\
  \bibinfo {pages} {244112} (\bibinfo {year} {2019})}\BibitemShut {NoStop}%
\bibitem [{\citenamefont {Sokolov}\ \emph {et~al.}(2020)\citenamefont
  {Sokolov}, \citenamefont {Barkoutsos}, \citenamefont {Ollitrault},
  \citenamefont {Greenberg}, \citenamefont {Rice}, \citenamefont {Pistoia},\
  and\ \citenamefont {Tavernelli}}]{Sokolov20}%
  \BibitemOpen
  \bibfield  {author} {\bibinfo {author} {\bibfnamefont {I.~O.}\ \bibnamefont
  {Sokolov}}, \bibinfo {author} {\bibfnamefont {P.~K.}\ \bibnamefont
  {Barkoutsos}}, \bibinfo {author} {\bibfnamefont {P.~J.}\ \bibnamefont
  {Ollitrault}}, \bibinfo {author} {\bibfnamefont {D.}~\bibnamefont
  {Greenberg}}, \bibinfo {author} {\bibfnamefont {J.}~\bibnamefont {Rice}},
  \bibinfo {author} {\bibfnamefont {M.}~\bibnamefont {Pistoia}}, \ and\
  \bibinfo {author} {\bibfnamefont {I.}~\bibnamefont {Tavernelli}},\ }\href
  {\doibase 10.1063/1.5141835} {\bibfield  {journal} {\bibinfo  {journal} {J.
  Chem. Phys.}\ }\textbf {\bibinfo {volume} {152}},\ \bibinfo {pages} {124107}
  (\bibinfo {year} {2020})}\BibitemShut {NoStop}%
\bibitem [{\citenamefont {Lee}\ \emph {et~al.}(2019)\citenamefont {Lee},
  \citenamefont {Huggins}, \citenamefont {Head-Gordon},\ and\ \citenamefont
  {Whaley}}]{Lee19}%
  \BibitemOpen
  \bibfield  {author} {\bibinfo {author} {\bibfnamefont {J.}~\bibnamefont
  {Lee}}, \bibinfo {author} {\bibfnamefont {W.~J.}\ \bibnamefont {Huggins}},
  \bibinfo {author} {\bibfnamefont {M.}~\bibnamefont {Head-Gordon}}, \ and\
  \bibinfo {author} {\bibfnamefont {K.~B.}\ \bibnamefont {Whaley}},\
  }\href@noop {} {\bibfield  {journal} {\bibinfo  {journal} {J. Chem. Theory
  Comput.}\ }\textbf {\bibinfo {volume} {15}},\ \bibinfo {pages} {311}
  (\bibinfo {year} {2019})}\BibitemShut {NoStop}%
\bibitem [{\citenamefont {Grimsley}\ \emph {et~al.}(2019)\citenamefont
  {Grimsley}, \citenamefont {Economou}, \citenamefont {Barnes},\ and\
  \citenamefont {Mayhall}}]{Grimsley19}%
  \BibitemOpen
  \bibfield  {author} {\bibinfo {author} {\bibfnamefont {H.~R.}\ \bibnamefont
  {Grimsley}}, \bibinfo {author} {\bibfnamefont {S.~E.}\ \bibnamefont
  {Economou}}, \bibinfo {author} {\bibfnamefont {E.}~\bibnamefont {Barnes}}, \
  and\ \bibinfo {author} {\bibfnamefont {N.~J.}\ \bibnamefont {Mayhall}},\
  }\href@noop {} {\bibfield  {journal} {\bibinfo  {journal} {Nat. Comm.}\
  }\textbf {\bibinfo {volume} {10}},\ \bibinfo {pages} {3007} (\bibinfo {year}
  {2019})}\BibitemShut {NoStop}%
\bibitem [{\citenamefont {Handy}\ \emph {et~al.}(1985)\citenamefont {Handy},
  \citenamefont {Knowles},\ and\ \citenamefont {Somasundram}}]{Handy85}%
  \BibitemOpen
  \bibfield  {author} {\bibinfo {author} {\bibfnamefont {N.}~\bibnamefont
  {Handy}}, \bibinfo {author} {\bibfnamefont {P.}~\bibnamefont {Knowles}}, \
  and\ \bibinfo {author} {\bibfnamefont {K.}~\bibnamefont {Somasundram}},\
  }\href@noop {} {\bibfield  {journal} {\bibinfo  {journal} {Theor. Chim.
  Acta}\ }\textbf {\bibinfo {volume} {68}},\ \bibinfo {pages} {87} (\bibinfo
  {year} {1985})}\BibitemShut {NoStop}%
\bibitem [{\citenamefont {Andrews}\ \emph {et~al.}(1991)\citenamefont
  {Andrews}, \citenamefont {Jayatilaka}, \citenamefont {Bone}, \citenamefont
  {Handy},\ and\ \citenamefont {Amos}}]{Andrews91}%
  \BibitemOpen
  \bibfield  {author} {\bibinfo {author} {\bibfnamefont {J.~S.}\ \bibnamefont
  {Andrews}}, \bibinfo {author} {\bibfnamefont {D.}~\bibnamefont {Jayatilaka}},
  \bibinfo {author} {\bibfnamefont {R.~G.}\ \bibnamefont {Bone}}, \bibinfo
  {author} {\bibfnamefont {N.~C.}\ \bibnamefont {Handy}}, \ and\ \bibinfo
  {author} {\bibfnamefont {R.~D.}\ \bibnamefont {Amos}},\ }\href {\doibase
  https://doi.org/10.1016/0009-2614(91)90405-X} {\bibfield  {journal} {\bibinfo
   {journal} {Chem. Phys. Lett.}\ }\textbf {\bibinfo {volume} {183}},\ \bibinfo
  {pages} {423 } (\bibinfo {year} {1991})}\BibitemShut {NoStop}%
\bibitem [{\citenamefont {Tsuchimochi}\ and\ \citenamefont
  {Scuseria}(2010)}]{Tsuchimochi10B}%
  \BibitemOpen
  \bibfield  {author} {\bibinfo {author} {\bibfnamefont {T.}~\bibnamefont
  {Tsuchimochi}}\ and\ \bibinfo {author} {\bibfnamefont {G.~E.}\ \bibnamefont
  {Scuseria}},\ }\href@noop {} {\bibfield  {journal} {\bibinfo  {journal} {J.
  Chem. Phys.}\ }\textbf {\bibinfo {volume} {133}},\ \bibinfo {pages} {141102}
  (\bibinfo {year} {2010})}\BibitemShut {NoStop}%
\bibitem [{\citenamefont {Paldus}(1977)}]{Paldus77}%
  \BibitemOpen
  \bibfield  {author} {\bibinfo {author} {\bibfnamefont {J.}~\bibnamefont
  {Paldus}},\ }\href@noop {} {\bibfield  {journal} {\bibinfo  {journal} {J.
  Chem. Phys.}\ ,\ \bibinfo {pages} {303}} (\bibinfo {year}
  {1977})}\BibitemShut {NoStop}%
\bibitem [{\citenamefont {Scuseria}\ \emph {et~al.}(1988)\citenamefont
  {Scuseria}, \citenamefont {Scheiner}, \citenamefont {Lee}, \citenamefont
  {Rice},\ and\ \citenamefont {{Schaefer, III}}}]{Scuseria87}%
  \BibitemOpen
  \bibfield  {author} {\bibinfo {author} {\bibfnamefont {G.~E.}\ \bibnamefont
  {Scuseria}}, \bibinfo {author} {\bibfnamefont {A.~C.}\ \bibnamefont
  {Scheiner}}, \bibinfo {author} {\bibfnamefont {T.~J.}\ \bibnamefont {Lee}},
  \bibinfo {author} {\bibfnamefont {J.~E.}\ \bibnamefont {Rice}}, \ and\
  \bibinfo {author} {\bibfnamefont {H.~F.}\ \bibnamefont {{Schaefer, III}}},\
  }\href@noop {} {\bibfield  {journal} {\bibinfo  {journal} {J. Chem. Phys.}\
  }\textbf {\bibinfo {volume} {86}},\ \bibinfo {pages} {2881} (\bibinfo {year}
  {1988})}\BibitemShut {NoStop}%
\bibitem [{\citenamefont {McClean}\ \emph {et~al.}(2016)\citenamefont
  {McClean}, \citenamefont {Romero}, \citenamefont {Babbush},\ and\
  \citenamefont {Aspuru-Guzik}}]{McClean16}%
  \BibitemOpen
  \bibfield  {author} {\bibinfo {author} {\bibfnamefont {J.~R.}\ \bibnamefont
  {McClean}}, \bibinfo {author} {\bibfnamefont {J.}~\bibnamefont {Romero}},
  \bibinfo {author} {\bibfnamefont {R.}~\bibnamefont {Babbush}}, \ and\
  \bibinfo {author} {\bibfnamefont {A.}~\bibnamefont {Aspuru-Guzik}},\
  }\href@noop {} {\bibfield  {journal} {\bibinfo  {journal} {New J. Phys.}\
  }\textbf {\bibinfo {volume} {18}},\ \bibinfo {pages} {023023} (\bibinfo
  {year} {2016})}\BibitemShut {NoStop}%
\bibitem [{\citenamefont {Yen}\ \emph {et~al.}(2019)\citenamefont {Yen},
  \citenamefont {Lang},\ and\ \citenamefont {Izmaylov}}]{Yen19}%
  \BibitemOpen
  \bibfield  {author} {\bibinfo {author} {\bibfnamefont {T.-C.}\ \bibnamefont
  {Yen}}, \bibinfo {author} {\bibfnamefont {R.~A.}\ \bibnamefont {Lang}}, \
  and\ \bibinfo {author} {\bibfnamefont {A.~F.}\ \bibnamefont {Izmaylov}},\
  }\href@noop {} {\bibfield  {journal} {\bibinfo  {journal} {J. Chem. Phys.}\
  }\textbf {\bibinfo {volume} {151}},\ \bibinfo {pages} {164111} (\bibinfo
  {year} {2019})}\BibitemShut {NoStop}%
\bibitem [{\citenamefont {Ryabinkin}\ \emph {et~al.}(2019)\citenamefont
  {Ryabinkin}, \citenamefont {Genin}, ,\ and\ \citenamefont
  {Izmaylov}}]{Ryabinkin19}%
  \BibitemOpen
  \bibfield  {author} {\bibinfo {author} {\bibfnamefont {I.~G.}\ \bibnamefont
  {Ryabinkin}}, \bibinfo {author} {\bibfnamefont {S.~N.}\ \bibnamefont
  {Genin}}, , \ and\ \bibinfo {author} {\bibfnamefont {A.~F.}\ \bibnamefont
  {Izmaylov}},\ }\href@noop {} {\bibfield  {journal} {\bibinfo  {journal} {J.
  Chem. Theory Comput.}\ ,\ \bibinfo {pages} {249}} (\bibinfo {year}
  {2019})}\BibitemShut {NoStop}%
\bibitem [{\citenamefont {Schlegel}(1986)}]{Schlegel86}%
  \BibitemOpen
  \bibfield  {author} {\bibinfo {author} {\bibfnamefont {H.~B.}\ \bibnamefont
  {Schlegel}},\ }\href@noop {} {\bibfield  {journal} {\bibinfo  {journal} {J.
  Chem. Phys.}\ }\textbf {\bibinfo {volume} {84}},\ \bibinfo {pages} {4530}
  (\bibinfo {year} {1986})}\BibitemShut {NoStop}%
\bibitem [{\citenamefont {Schlegel}(1988)}]{Schlegel88}%
  \BibitemOpen
  \bibfield  {author} {\bibinfo {author} {\bibfnamefont {H.~B.}\ \bibnamefont
  {Schlegel}},\ }\href@noop {} {\bibfield  {journal} {\bibinfo  {journal} {J.
  Phys. Chem.}\ }\textbf {\bibinfo {volume} {92}},\ \bibinfo {pages} {3075}
  (\bibinfo {year} {1988})}\BibitemShut {NoStop}%
\bibitem [{\citenamefont {Knowles}\ and\ \citenamefont
  {Handy}(1988)}]{Knowles88B}%
  \BibitemOpen
  \bibfield  {author} {\bibinfo {author} {\bibfnamefont {P.~J.}\ \bibnamefont
  {Knowles}}\ and\ \bibinfo {author} {\bibfnamefont {N.~C.}\ \bibnamefont
  {Handy}},\ }\href@noop {} {\bibfield  {journal} {\bibinfo  {journal} {J.
  Chem. Phys.}\ }\textbf {\bibinfo {volume} {88}},\ \bibinfo {pages} {6991}
  (\bibinfo {year} {1988})}\BibitemShut {NoStop}%
\bibitem [{\citenamefont {L$\rm\ddot{o}$wdin}(1955)}]{Lowdin55B}%
  \BibitemOpen
  \bibfield  {author} {\bibinfo {author} {\bibfnamefont {P.-O.}\ \bibnamefont
  {L$\rm\ddot{o}$wdin}},\ }\href@noop {} {\bibfield  {journal} {\bibinfo
  {journal} {Phys. Rev.}\ }\textbf {\bibinfo {volume} {97}},\ \bibinfo {pages}
  {1509} (\bibinfo {year} {1955})}\BibitemShut {NoStop}%
\bibitem [{\citenamefont {Percus}\ and\ \citenamefont
  {Rotenberg}(1962)}]{Percus62}%
  \BibitemOpen
  \bibfield  {author} {\bibinfo {author} {\bibfnamefont {J.~K.}\ \bibnamefont
  {Percus}}\ and\ \bibinfo {author} {\bibfnamefont {A.}~\bibnamefont
  {Rotenberg}},\ }\href@noop {} {\bibfield  {journal} {\bibinfo  {journal} {J.
  Math. Phys.}\ }\textbf {\bibinfo {volume} {3}},\ \bibinfo {pages} {928}
  (\bibinfo {year} {1962})}\BibitemShut {NoStop}%
\bibitem [{\citenamefont {Scuseria}\ \emph {et~al.}(2011)\citenamefont
  {Scuseria}, \citenamefont {Jim\'enez-Hoyos}, \citenamefont {Henderson},
  \citenamefont {Samanta},\ and\ \citenamefont {Ellis}}]{Scuseria11}%
  \BibitemOpen
  \bibfield  {author} {\bibinfo {author} {\bibfnamefont {G.~E.}\ \bibnamefont
  {Scuseria}}, \bibinfo {author} {\bibfnamefont {C.~A.}\ \bibnamefont
  {Jim\'enez-Hoyos}}, \bibinfo {author} {\bibfnamefont {T.~M.}\ \bibnamefont
  {Henderson}}, \bibinfo {author} {\bibfnamefont {K.}~\bibnamefont {Samanta}},
  \ and\ \bibinfo {author} {\bibfnamefont {J.~K.}\ \bibnamefont {Ellis}},\
  }\href@noop {} {\bibfield  {journal} {\bibinfo  {journal} {J. Chem. Phys.}\
  }\textbf {\bibinfo {volume} {135}},\ \bibinfo {pages} {124108} (\bibinfo
  {year} {2011})}\BibitemShut {NoStop}%
\bibitem [{\citenamefont {Jim\'enez-Hoyos}\ \emph {et~al.}(2012)\citenamefont
  {Jim\'enez-Hoyos}, \citenamefont {Henderson}, \citenamefont {Tsuchimochi},\
  and\ \citenamefont {Scuseria}}]{Jimenez12}%
  \BibitemOpen
  \bibfield  {author} {\bibinfo {author} {\bibfnamefont {C.~A.}\ \bibnamefont
  {Jim\'enez-Hoyos}}, \bibinfo {author} {\bibfnamefont {T.~M.}\ \bibnamefont
  {Henderson}}, \bibinfo {author} {\bibfnamefont {T.}~\bibnamefont
  {Tsuchimochi}}, \ and\ \bibinfo {author} {\bibfnamefont {G.~E.}\ \bibnamefont
  {Scuseria}},\ }\href@noop {} {\bibfield  {journal} {\bibinfo  {journal} {J.
  Chem. Phys.}\ }\textbf {\bibinfo {volume} {136}},\ \bibinfo {pages} {164109}
  (\bibinfo {year} {2012})}\BibitemShut {NoStop}%
\bibitem [{\citenamefont {Lestrange}\ \emph {et~al.}(2018)\citenamefont
  {Lestrange}, \citenamefont {Williams-Young}, \citenamefont {Petrone},
  \citenamefont {Jim\'enez-Hoyos},\ and\ \citenamefont {Li}}]{Lestrange18}%
  \BibitemOpen
  \bibfield  {author} {\bibinfo {author} {\bibfnamefont {P.~J.}\ \bibnamefont
  {Lestrange}}, \bibinfo {author} {\bibfnamefont {D.~B.}\ \bibnamefont
  {Williams-Young}}, \bibinfo {author} {\bibfnamefont {A.}~\bibnamefont
  {Petrone}}, \bibinfo {author} {\bibfnamefont {C.~A.}\ \bibnamefont
  {Jim\'enez-Hoyos}}, \ and\ \bibinfo {author} {\bibfnamefont {X.}~\bibnamefont
  {Li}},\ }\href {\doibase 10.1021/acs.jctc.7b00832} {\bibfield  {journal}
  {\bibinfo  {journal} {J. Chem. Theory Comput.}\ }\textbf {\bibinfo {volume}
  {14}},\ \bibinfo {pages} {588} (\bibinfo {year} {2018})}\BibitemShut
  {NoStop}%
\bibitem [{\citenamefont {Qiu}\ \emph {et~al.}(2017{\natexlab{a}})\citenamefont
  {Qiu}, \citenamefont {Henderson}, \citenamefont {Zhao},\ and\ \citenamefont
  {Scuseria}}]{Qiu17B}%
  \BibitemOpen
  \bibfield  {author} {\bibinfo {author} {\bibfnamefont {Y.}~\bibnamefont
  {Qiu}}, \bibinfo {author} {\bibfnamefont {T.~M.}\ \bibnamefont {Henderson}},
  \bibinfo {author} {\bibfnamefont {J.}~\bibnamefont {Zhao}}, \ and\ \bibinfo
  {author} {\bibfnamefont {G.~E.}\ \bibnamefont {Scuseria}},\ }\href@noop {}
  {\bibfield  {journal} {\bibinfo  {journal} {J. Chem. Phys.}\ }\textbf
  {\bibinfo {volume} {147}},\ \bibinfo {pages} {064111} (\bibinfo {year}
  {2017}{\natexlab{a}})}\BibitemShut {NoStop}%
\bibitem [{\citenamefont {Qiu}\ \emph {et~al.}(2018)\citenamefont {Qiu},
  \citenamefont {Henderson}, \citenamefont {Zhao},\ and\ \citenamefont
  {Scuseria}}]{Qiu18}%
  \BibitemOpen
  \bibfield  {author} {\bibinfo {author} {\bibfnamefont {Y.}~\bibnamefont
  {Qiu}}, \bibinfo {author} {\bibfnamefont {T.~M.}\ \bibnamefont {Henderson}},
  \bibinfo {author} {\bibfnamefont {J.}~\bibnamefont {Zhao}}, \ and\ \bibinfo
  {author} {\bibfnamefont {G.~E.}\ \bibnamefont {Scuseria}},\ }\href@noop {}
  {\bibfield  {journal} {\bibinfo  {journal} {J. Chem. Phys.}\ }\textbf
  {\bibinfo {volume} {149}},\ \bibinfo {pages} {064111} (\bibinfo {year}
  {2018})}\BibitemShut {NoStop}%
\bibitem [{\citenamefont {Tsuchimochi}\ and\ \citenamefont
  {Ten-no}(2018)}]{Tsuchimochi18}%
  \BibitemOpen
  \bibfield  {author} {\bibinfo {author} {\bibfnamefont {T.}~\bibnamefont
  {Tsuchimochi}}\ and\ \bibinfo {author} {\bibfnamefont {S.~L.}\ \bibnamefont
  {Ten-no}},\ }\href@noop {} {\bibfield  {journal} {\bibinfo  {journal} {J.
  Chem. Phys.}\ }\textbf {\bibinfo {volume} {149}},\ \bibinfo {pages} {044109}
  (\bibinfo {year} {2018})}\BibitemShut {NoStop}%
\bibitem [{\citenamefont {Tsuchimochi}\ and\ \citenamefont
  {Ten-no}(2019)}]{Tsuchimochi19A}%
  \BibitemOpen
  \bibfield  {author} {\bibinfo {author} {\bibfnamefont {T.}~\bibnamefont
  {Tsuchimochi}}\ and\ \bibinfo {author} {\bibfnamefont {S.~L.}\ \bibnamefont
  {Ten-no}},\ }\href@noop {} {\bibfield  {journal} {\bibinfo  {journal} {J.
  Comput. Chem.}\ }\textbf {\bibinfo {volume} {40}},\ \bibinfo {pages} {267}
  (\bibinfo {year} {2019})}\BibitemShut {NoStop}%
\bibitem [{\citenamefont {Garcia-Escartin}\ and\ \citenamefont
  {Chamorro-Posada}(2013)}]{Garcia13}%
  \BibitemOpen
  \bibfield  {author} {\bibinfo {author} {\bibfnamefont {J.~C.}\ \bibnamefont
  {Garcia-Escartin}}\ and\ \bibinfo {author} {\bibfnamefont {P.}~\bibnamefont
  {Chamorro-Posada}},\ }\href {\doibase 10.1103/PhysRevA.87.052330} {\bibfield
  {journal} {\bibinfo  {journal} {Phys. Rev. A}\ }\textbf {\bibinfo {volume}
  {87}},\ \bibinfo {pages} {052330} (\bibinfo {year} {2013})}\BibitemShut
  {NoStop}%
\bibitem [{\citenamefont {Mitarai}\ and\ \citenamefont
  {Fujii}(2019)}]{Mitarai19}%
  \BibitemOpen
  \bibfield  {author} {\bibinfo {author} {\bibfnamefont {K.}~\bibnamefont
  {Mitarai}}\ and\ \bibinfo {author} {\bibfnamefont {K.}~\bibnamefont
  {Fujii}},\ }\href {\doibase 10.1103/PhysRevResearch.1.013006} {\bibfield
  {journal} {\bibinfo  {journal} {Phys. Rev. Research}\ }\textbf {\bibinfo
  {volume} {1}},\ \bibinfo {pages} {013006} (\bibinfo {year}
  {2019})}\BibitemShut {NoStop}%
\bibitem [{\citenamefont {Izmaylov}\ \emph {et~al.}(2020)\citenamefont
  {Izmaylov}, \citenamefont {Yen}, \citenamefont {Lang},\ and\ \citenamefont
  {Verteletskyi}}]{Izmaylov20}%
  \BibitemOpen
  \bibfield  {author} {\bibinfo {author} {\bibfnamefont {A.~F.}\ \bibnamefont
  {Izmaylov}}, \bibinfo {author} {\bibfnamefont {T.-C.}\ \bibnamefont {Yen}},
  \bibinfo {author} {\bibfnamefont {R.~A.}\ \bibnamefont {Lang}}, \ and\
  \bibinfo {author} {\bibfnamefont {V.}~\bibnamefont {Verteletskyi}},\
  }\href@noop {} {\bibfield  {journal} {\bibinfo  {journal} {J. Chem. Theory
  Comput.}\ ,\ \bibinfo {pages} {190}} (\bibinfo {year} {2020})}\BibitemShut
  {NoStop}%
\bibitem [{\citenamefont {Parrish}\ \emph {et~al.}(2019)\citenamefont
  {Parrish}, \citenamefont {Hohenstein}, \citenamefont {McMahon},\ and\
  \citenamefont {Mart\'{\i}nez}}]{Parrish19}%
  \BibitemOpen
  \bibfield  {author} {\bibinfo {author} {\bibfnamefont {R.~M.}\ \bibnamefont
  {Parrish}}, \bibinfo {author} {\bibfnamefont {E.~G.}\ \bibnamefont
  {Hohenstein}}, \bibinfo {author} {\bibfnamefont {P.~L.}\ \bibnamefont
  {McMahon}}, \ and\ \bibinfo {author} {\bibfnamefont {T.~J.}\ \bibnamefont
  {Mart\'{\i}nez}},\ }\href {\doibase 10.1103/PhysRevLett.122.230401}
  {\bibfield  {journal} {\bibinfo  {journal} {Phys. Rev. Lett.}\ }\textbf
  {\bibinfo {volume} {122}},\ \bibinfo {pages} {230401} (\bibinfo {year}
  {2019})}\BibitemShut {NoStop}%
\bibitem [{\citenamefont {Huggins}\ \emph {et~al.}(2020)\citenamefont
  {Huggins}, \citenamefont {Lee}, \citenamefont {Baek}, \citenamefont
  {O'Gorman},\ and\ \citenamefont {Whaley}}]{Huggins20}%
  \BibitemOpen
  \bibfield  {author} {\bibinfo {author} {\bibfnamefont {W.~J.}\ \bibnamefont
  {Huggins}}, \bibinfo {author} {\bibfnamefont {J.}~\bibnamefont {Lee}},
  \bibinfo {author} {\bibfnamefont {U.}~\bibnamefont {Baek}}, \bibinfo {author}
  {\bibfnamefont {B.}~\bibnamefont {O'Gorman}}, \ and\ \bibinfo {author}
  {\bibfnamefont {K.~B.}\ \bibnamefont {Whaley}},\ }\href
  {http://iopscience.iop.org/10.1088/1367-2630/ab867b} {\bibfield  {journal}
  {\bibinfo  {journal} {New J. Phys.}\ } (\bibinfo {year} {2020})},\ \bibinfo
  {note} {in press}\BibitemShut {NoStop}%
\bibitem [{\citenamefont {Kivlichan}\ \emph {et~al.}(2018)\citenamefont
  {Kivlichan}, \citenamefont {McClean}, \citenamefont {Wiebe}, \citenamefont
  {Gidney}, \citenamefont {Aspuru-Guzik}, \citenamefont {Chan},\ and\
  \citenamefont {Babbush}}]{Kivlichan18}%
  \BibitemOpen
  \bibfield  {author} {\bibinfo {author} {\bibfnamefont {I.~D.}\ \bibnamefont
  {Kivlichan}}, \bibinfo {author} {\bibfnamefont {J.}~\bibnamefont {McClean}},
  \bibinfo {author} {\bibfnamefont {N.}~\bibnamefont {Wiebe}}, \bibinfo
  {author} {\bibfnamefont {C.}~\bibnamefont {Gidney}}, \bibinfo {author}
  {\bibfnamefont {A.}~\bibnamefont {Aspuru-Guzik}}, \bibinfo {author}
  {\bibfnamefont {G.~K.-L.}\ \bibnamefont {Chan}}, \ and\ \bibinfo {author}
  {\bibfnamefont {R.}~\bibnamefont {Babbush}},\ }\href {\doibase
  10.1103/PhysRevLett.120.110501} {\bibfield  {journal} {\bibinfo  {journal}
  {Phys. Rev. Lett.}\ }\textbf {\bibinfo {volume} {120}},\ \bibinfo {pages}
  {110501} (\bibinfo {year} {2018})}\BibitemShut {NoStop}%
\bibitem [{\citenamefont {Tsuchimochi}(2015)}]{Tsuchimochi15C}%
  \BibitemOpen
  \bibfield  {author} {\bibinfo {author} {\bibfnamefont {T.}~\bibnamefont
  {Tsuchimochi}},\ }\href@noop {} {\bibfield  {journal} {\bibinfo  {journal}
  {J. Chem. Phys.}\ }\textbf {\bibinfo {volume} {143}},\ \bibinfo {pages}
  {144114} (\bibinfo {year} {2015})}\BibitemShut {NoStop}%
\bibitem [{\citenamefont {Tsuchimochi}\ and\ \citenamefont
  {Ten-no}(2016)}]{Tsuchimochi16A}%
  \BibitemOpen
  \bibfield  {author} {\bibinfo {author} {\bibfnamefont {T.}~\bibnamefont
  {Tsuchimochi}}\ and\ \bibinfo {author} {\bibfnamefont {S.}~\bibnamefont
  {Ten-no}},\ }\href@noop {} {\bibfield  {journal} {\bibinfo  {journal} {J.
  Chem. Phys.}\ }\textbf {\bibinfo {volume} {144}},\ \bibinfo {pages} {011101}
  (\bibinfo {year} {2016})}\BibitemShut {NoStop}%
\bibitem [{\citenamefont {Wecker}\ \emph {et~al.}(2015)\citenamefont {Wecker},
  \citenamefont {Hastings}, \citenamefont {Wiebe}, \citenamefont {Clark},
  \citenamefont {Nayak},\ and\ \citenamefont {Troyer}}]{Wecker15}%
  \BibitemOpen
  \bibfield  {author} {\bibinfo {author} {\bibfnamefont {D.}~\bibnamefont
  {Wecker}}, \bibinfo {author} {\bibfnamefont {M.~B.}\ \bibnamefont
  {Hastings}}, \bibinfo {author} {\bibfnamefont {N.}~\bibnamefont {Wiebe}},
  \bibinfo {author} {\bibfnamefont {B.~K.}\ \bibnamefont {Clark}}, \bibinfo
  {author} {\bibfnamefont {C.}~\bibnamefont {Nayak}}, \ and\ \bibinfo {author}
  {\bibfnamefont {M.}~\bibnamefont {Troyer}},\ }\href {\doibase
  10.1103/PhysRevA.92.062318} {\bibfield  {journal} {\bibinfo  {journal} {Phys.
  Rev. A}\ }\textbf {\bibinfo {volume} {92}},\ \bibinfo {pages} {062318}
  (\bibinfo {year} {2015})}\BibitemShut {NoStop}%
\bibitem [{\citenamefont {Thouless}(1960)}]{Thouless60}%
  \BibitemOpen
  \bibfield  {author} {\bibinfo {author} {\bibfnamefont {D.}~\bibnamefont
  {Thouless}},\ }\href@noop {} {\bibfield  {journal} {\bibinfo  {journal}
  {Nucl. Phys.}\ }\textbf {\bibinfo {volume} {21}},\ \bibinfo {pages} {225}
  (\bibinfo {year} {1960})}\BibitemShut {NoStop}%
\bibitem [{\citenamefont {Sherrill}\ \emph {et~al.}(1998)\citenamefont
  {Sherrill}, \citenamefont {Krylov}, \citenamefont {Byrd},\ and\ \citenamefont
  {Head-Gordon}}]{Sherrill98}%
  \BibitemOpen
  \bibfield  {author} {\bibinfo {author} {\bibfnamefont {C.~D.}\ \bibnamefont
  {Sherrill}}, \bibinfo {author} {\bibfnamefont {A.~I.}\ \bibnamefont
  {Krylov}}, \bibinfo {author} {\bibfnamefont {E.~F.~C.}\ \bibnamefont {Byrd}},
  \ and\ \bibinfo {author} {\bibfnamefont {M.}~\bibnamefont {Head-Gordon}},\
  }\href@noop {} {\bibfield  {journal} {\bibinfo  {journal} {J. Chem. Phys.}\
  }\textbf {\bibinfo {volume} {109}},\ \bibinfo {pages} {4171} (\bibinfo {year}
  {1998})}\BibitemShut {NoStop}%
\bibitem [{\citenamefont {Helgaker}\ \emph {et~al.}(2000)\citenamefont
  {Helgaker}, \citenamefont {J{\o}rgensen},\ and\ \citenamefont
  {Olsen}}]{Helgaker00}%
  \BibitemOpen
  \bibfield  {author} {\bibinfo {author} {\bibfnamefont {T.}~\bibnamefont
  {Helgaker}}, \bibinfo {author} {\bibfnamefont {P.}~\bibnamefont
  {J{\o}rgensen}}, \ and\ \bibinfo {author} {\bibfnamefont {J.}~\bibnamefont
  {Olsen}},\ }\href@noop {} {\emph {\bibinfo {title} {Molecular
  Electronic-Structure Theory}}}\ (\bibinfo  {publisher} {Wiley},\ \bibinfo
  {address} {Chichester, U.~K.},\ \bibinfo {year} {2000})\BibitemShut {NoStop}%
\bibitem [{\citenamefont {Mizukami}\ \emph {et~al.}(2019)\citenamefont
  {Mizukami}, \citenamefont {Mitarai}, \citenamefont {Nakagawa}, \citenamefont
  {Yamamoto}, \citenamefont {Yan},\ and\ \citenamefont
  {ya~Ohnishi}}]{Mizukami19}%
  \BibitemOpen
  \bibfield  {author} {\bibinfo {author} {\bibfnamefont {W.}~\bibnamefont
  {Mizukami}}, \bibinfo {author} {\bibfnamefont {K.}~\bibnamefont {Mitarai}},
  \bibinfo {author} {\bibfnamefont {Y.~O.}\ \bibnamefont {Nakagawa}}, \bibinfo
  {author} {\bibfnamefont {T.}~\bibnamefont {Yamamoto}}, \bibinfo {author}
  {\bibfnamefont {T.}~\bibnamefont {Yan}}, \ and\ \bibinfo {author}
  {\bibfnamefont {Y.}~\bibnamefont {ya~Ohnishi}},\ }\href@noop {} {\  (\bibinfo
  {year} {2019})},\ \Eprint {http://arxiv.org/abs/1910.11526} {arXiv:1910.11526
  [cond-mat]} \BibitemShut {NoStop}%
\bibitem [{com()}]{comment}%
  \BibitemOpen
  \href@noop {} {}\bibinfo {note} {One can further introduce $\kappa_{q}^{\bar
  p}$ for the rotations between $\alpha$ and $\beta$ spin-orbitals to obtain a
  generalized Hartree-Fock picture.}\BibitemShut {Stop}%
\bibitem [{\citenamefont {Chen}\ and\ \citenamefont {Hoffmann}(2012)}]{Chen12}%
  \BibitemOpen
  \bibfield  {author} {\bibinfo {author} {\bibfnamefont {Z.}~\bibnamefont
  {Chen}}\ and\ \bibinfo {author} {\bibfnamefont {M.~R.}\ \bibnamefont
  {Hoffmann}},\ }\href@noop {} {\bibfield  {journal} {\bibinfo  {journal} {J.
  Chem. Phys.}\ }\textbf {\bibinfo {volume} {137}},\ \bibinfo {pages} {014108}
  (\bibinfo {year} {2012})}\BibitemShut {NoStop}%
\bibitem [{\citenamefont {Sun}\ \emph {et~al.}(2018)\citenamefont {Sun},
  \citenamefont {Berkelbach}, \citenamefont {Blunt}, \citenamefont {Booth},
  \citenamefont {Guo}, \citenamefont {Li}, \citenamefont {Liu}, \citenamefont
  {McClain}, \citenamefont {Sayfutyarova}, \citenamefont {Sharma},
  \citenamefont {Wouters},\ and\ \citenamefont {Chan}}]{pyscf}%
  \BibitemOpen
  \bibfield  {author} {\bibinfo {author} {\bibfnamefont {Q.}~\bibnamefont
  {Sun}}, \bibinfo {author} {\bibfnamefont {T.~C.}\ \bibnamefont {Berkelbach}},
  \bibinfo {author} {\bibfnamefont {N.~S.}\ \bibnamefont {Blunt}}, \bibinfo
  {author} {\bibfnamefont {G.~H.}\ \bibnamefont {Booth}}, \bibinfo {author}
  {\bibfnamefont {S.}~\bibnamefont {Guo}}, \bibinfo {author} {\bibfnamefont
  {Z.}~\bibnamefont {Li}}, \bibinfo {author} {\bibfnamefont {J.}~\bibnamefont
  {Liu}}, \bibinfo {author} {\bibfnamefont {J.~D.}\ \bibnamefont {McClain}},
  \bibinfo {author} {\bibfnamefont {E.~R.}\ \bibnamefont {Sayfutyarova}},
  \bibinfo {author} {\bibfnamefont {S.}~\bibnamefont {Sharma}}, \bibinfo
  {author} {\bibfnamefont {S.}~\bibnamefont {Wouters}}, \ and\ \bibinfo
  {author} {\bibfnamefont {G.~K.-L.}\ \bibnamefont {Chan}},\ }\href {\doibase
  10.1002/wcms.1340} {\bibfield  {journal} {\bibinfo  {journal} {WIREs
  Computational Molecular Science}\ }\textbf {\bibinfo {volume} {8}},\ \bibinfo
  {pages} {e1340} (\bibinfo {year} {2018})}\BibitemShut {NoStop}%
\bibitem [{\citenamefont {McClean}\ \emph {et~al.}(2020)\citenamefont
  {McClean}, \citenamefont {Rubin}, \citenamefont {Sung}, \citenamefont
  {Kivlichan}, \citenamefont {Bonet-Monroig}, \citenamefont {Cao},
  \citenamefont {Dai}, \citenamefont {Fried}, \citenamefont {Gidney},
  \citenamefont {Gimby}, \citenamefont {Gokhale}, \citenamefont {H\"aner},
  \citenamefont {Hardikar}, \citenamefont {Havl{\'{\i}}{\v{c}}ek},
  \citenamefont {Higgott}, \citenamefont {Huang}, \citenamefont {Izaac},
  \citenamefont {Jiang}, \citenamefont {Liu}, \citenamefont {McArdle},
  \citenamefont {Neeley}, \citenamefont {O'Brien}, \citenamefont {O'Gorman},
  \citenamefont {Ozfidan}, \citenamefont {Radin}, \citenamefont {Romero},
  \citenamefont {Sawaya}, \citenamefont {Senjean}, \citenamefont {Setia},
  \citenamefont {Sim}, \citenamefont {Steiger}, \citenamefont {Steudtner},
  \citenamefont {Sun}, \citenamefont {Sun}, \citenamefont {Wang}, \citenamefont
  {Zhang},\ and\ \citenamefont {Babbush}}]{openfermion}%
  \BibitemOpen
  \bibfield  {author} {\bibinfo {author} {\bibfnamefont {J.~R.}\ \bibnamefont
  {McClean}}, \bibinfo {author} {\bibfnamefont {N.~C.}\ \bibnamefont {Rubin}},
  \bibinfo {author} {\bibfnamefont {K.~J.}\ \bibnamefont {Sung}}, \bibinfo
  {author} {\bibfnamefont {I.~D.}\ \bibnamefont {Kivlichan}}, \bibinfo {author}
  {\bibfnamefont {X.}~\bibnamefont {Bonet-Monroig}}, \bibinfo {author}
  {\bibfnamefont {Y.}~\bibnamefont {Cao}}, \bibinfo {author} {\bibfnamefont
  {C.}~\bibnamefont {Dai}}, \bibinfo {author} {\bibfnamefont {E.~S.}\
  \bibnamefont {Fried}}, \bibinfo {author} {\bibfnamefont {C.}~\bibnamefont
  {Gidney}}, \bibinfo {author} {\bibfnamefont {B.}~\bibnamefont {Gimby}},
  \bibinfo {author} {\bibfnamefont {P.}~\bibnamefont {Gokhale}}, \bibinfo
  {author} {\bibfnamefont {T.}~\bibnamefont {H\"aner}}, \bibinfo {author}
  {\bibfnamefont {T.}~\bibnamefont {Hardikar}}, \bibinfo {author}
  {\bibfnamefont {V.}~\bibnamefont {Havl{\'{\i}}{\v{c}}ek}}, \bibinfo {author}
  {\bibfnamefont {O.}~\bibnamefont {Higgott}}, \bibinfo {author} {\bibfnamefont
  {C.}~\bibnamefont {Huang}}, \bibinfo {author} {\bibfnamefont
  {J.}~\bibnamefont {Izaac}}, \bibinfo {author} {\bibfnamefont
  {Z.}~\bibnamefont {Jiang}}, \bibinfo {author} {\bibfnamefont
  {X.}~\bibnamefont {Liu}}, \bibinfo {author} {\bibfnamefont {S.}~\bibnamefont
  {McArdle}}, \bibinfo {author} {\bibfnamefont {M.}~\bibnamefont {Neeley}},
  \bibinfo {author} {\bibfnamefont {T.}~\bibnamefont {O'Brien}}, \bibinfo
  {author} {\bibfnamefont {B.}~\bibnamefont {O'Gorman}}, \bibinfo {author}
  {\bibfnamefont {I.}~\bibnamefont {Ozfidan}}, \bibinfo {author} {\bibfnamefont
  {M.~D.}\ \bibnamefont {Radin}}, \bibinfo {author} {\bibfnamefont
  {J.}~\bibnamefont {Romero}}, \bibinfo {author} {\bibfnamefont {N.~P.~D.}\
  \bibnamefont {Sawaya}}, \bibinfo {author} {\bibfnamefont {B.}~\bibnamefont
  {Senjean}}, \bibinfo {author} {\bibfnamefont {K.}~\bibnamefont {Setia}},
  \bibinfo {author} {\bibfnamefont {S.}~\bibnamefont {Sim}}, \bibinfo {author}
  {\bibfnamefont {D.~S.}\ \bibnamefont {Steiger}}, \bibinfo {author}
  {\bibfnamefont {M.}~\bibnamefont {Steudtner}}, \bibinfo {author}
  {\bibfnamefont {Q.}~\bibnamefont {Sun}}, \bibinfo {author} {\bibfnamefont
  {W.}~\bibnamefont {Sun}}, \bibinfo {author} {\bibfnamefont {D.}~\bibnamefont
  {Wang}}, \bibinfo {author} {\bibfnamefont {F.}~\bibnamefont {Zhang}}, \ and\
  \bibinfo {author} {\bibfnamefont {R.}~\bibnamefont {Babbush}},\ }\href
  {\doibase 10.1088/2058-9565/ab8ebc} {\bibfield  {journal} {\bibinfo
  {journal} {Quantum Science and Technology}\ }\textbf {\bibinfo {volume}
  {5}},\ \bibinfo {pages} {034014} (\bibinfo {year} {2020})}\BibitemShut
  {NoStop}%
\bibitem [{Qul(2018)}]{Qulacs}%
  \BibitemOpen
  \href@noop {} {\enquote {\bibinfo {title} {{Qulacs}},}\ } (\bibinfo {year}
  {2018}),\ \Eprint {http://arxiv.org/abs/https://github.com/qulacs/qulacs}
  {https://github.com/qulacs/qulacs} \BibitemShut {NoStop}%
\bibitem [{\citenamefont {{Virtanen}}\ \emph {et~al.}(2020)\citenamefont
  {{Virtanen}}, \citenamefont {{Gommers}}, \citenamefont {{Oliphant}},
  \citenamefont {{Haberland}}, \citenamefont {{Reddy}}, \citenamefont
  {{Cournapeau}}, \citenamefont {{Burovski}}, \citenamefont {{Peterson}},
  \citenamefont {{Weckesser}}, \citenamefont {{Bright}}, \citenamefont {{van
  der Walt}}, \citenamefont {{Brett}}, \citenamefont {{Wilson}}, \citenamefont
  {{Jarrod Millman}}, \citenamefont {{Mayorov}}, \citenamefont {{Nelson}},
  \citenamefont {{Jones}}, \citenamefont {{Kern}}, \citenamefont {{Larson}},
  \citenamefont {{Carey}}, \citenamefont {{Polat}}, \citenamefont {{Feng}},
  \citenamefont {{Moore}}, \citenamefont {{Vand erPlas}}, \citenamefont
  {{Laxalde}}, \citenamefont {{Perktold}}, \citenamefont {{Cimrman}},
  \citenamefont {{Henriksen}}, \citenamefont {{Quintero}}, \citenamefont
  {{Harris}}, \citenamefont {{Archibald}}, \citenamefont {{Ribeiro}},
  \citenamefont {{Pedregosa}}, \citenamefont {{van Mulbregt}},\ and\
  \citenamefont {{SciPy 1.0 Contributors}}}]{scipy}%
  \BibitemOpen
  \bibfield  {author} {\bibinfo {author} {\bibfnamefont {P.}~\bibnamefont
  {{Virtanen}}}, \bibinfo {author} {\bibfnamefont {R.}~\bibnamefont
  {{Gommers}}}, \bibinfo {author} {\bibfnamefont {T.~E.}\ \bibnamefont
  {{Oliphant}}}, \bibinfo {author} {\bibfnamefont {M.}~\bibnamefont
  {{Haberland}}}, \bibinfo {author} {\bibfnamefont {T.}~\bibnamefont
  {{Reddy}}}, \bibinfo {author} {\bibfnamefont {D.}~\bibnamefont
  {{Cournapeau}}}, \bibinfo {author} {\bibfnamefont {E.}~\bibnamefont
  {{Burovski}}}, \bibinfo {author} {\bibfnamefont {P.}~\bibnamefont
  {{Peterson}}}, \bibinfo {author} {\bibfnamefont {W.}~\bibnamefont
  {{Weckesser}}}, \bibinfo {author} {\bibfnamefont {J.}~\bibnamefont
  {{Bright}}}, \bibinfo {author} {\bibfnamefont {S.~J.}\ \bibnamefont {{van der
  Walt}}}, \bibinfo {author} {\bibfnamefont {M.}~\bibnamefont {{Brett}}},
  \bibinfo {author} {\bibfnamefont {J.}~\bibnamefont {{Wilson}}}, \bibinfo
  {author} {\bibfnamefont {K.}~\bibnamefont {{Jarrod Millman}}}, \bibinfo
  {author} {\bibfnamefont {N.}~\bibnamefont {{Mayorov}}}, \bibinfo {author}
  {\bibfnamefont {A.~R.~J.}\ \bibnamefont {{Nelson}}}, \bibinfo {author}
  {\bibfnamefont {E.}~\bibnamefont {{Jones}}}, \bibinfo {author} {\bibfnamefont
  {R.}~\bibnamefont {{Kern}}}, \bibinfo {author} {\bibfnamefont
  {E.}~\bibnamefont {{Larson}}}, \bibinfo {author} {\bibfnamefont
  {C.}~\bibnamefont {{Carey}}}, \bibinfo {author} {\bibfnamefont
  {{\.I}.}~\bibnamefont {{Polat}}}, \bibinfo {author} {\bibfnamefont
  {Y.}~\bibnamefont {{Feng}}}, \bibinfo {author} {\bibfnamefont {E.~W.}\
  \bibnamefont {{Moore}}}, \bibinfo {author} {\bibfnamefont {J.}~\bibnamefont
  {{Vand erPlas}}}, \bibinfo {author} {\bibfnamefont {D.}~\bibnamefont
  {{Laxalde}}}, \bibinfo {author} {\bibfnamefont {J.}~\bibnamefont
  {{Perktold}}}, \bibinfo {author} {\bibfnamefont {R.}~\bibnamefont
  {{Cimrman}}}, \bibinfo {author} {\bibfnamefont {I.}~\bibnamefont
  {{Henriksen}}}, \bibinfo {author} {\bibfnamefont {E.~A.}\ \bibnamefont
  {{Quintero}}}, \bibinfo {author} {\bibfnamefont {C.~R.}\ \bibnamefont
  {{Harris}}}, \bibinfo {author} {\bibfnamefont {A.~M.}\ \bibnamefont
  {{Archibald}}}, \bibinfo {author} {\bibfnamefont {A.~H.}\ \bibnamefont
  {{Ribeiro}}}, \bibinfo {author} {\bibfnamefont {F.}~\bibnamefont
  {{Pedregosa}}}, \bibinfo {author} {\bibfnamefont {P.}~\bibnamefont {{van
  Mulbregt}}}, \ and\ \bibinfo {author} {\bibnamefont {{SciPy 1.0
  Contributors}}},\ }\href {\doibase https://doi.org/10.1038/s41592-019-0686-2}
  {\bibfield  {journal} {\bibinfo  {journal} {Nat. Methods}\ }\textbf {\bibinfo
  {volume} {17}},\ \bibinfo {pages} {261} (\bibinfo {year} {2020})}\BibitemShut
  {NoStop}%
\bibitem [{\citenamefont {Qiu}\ \emph {et~al.}(2017{\natexlab{b}})\citenamefont
  {Qiu}, \citenamefont {Henderson},\ and\ \citenamefont {Scuseria}}]{Qiu17A}%
  \BibitemOpen
  \bibfield  {author} {\bibinfo {author} {\bibfnamefont {Y.}~\bibnamefont
  {Qiu}}, \bibinfo {author} {\bibfnamefont {T.~M.}\ \bibnamefont {Henderson}},
  \ and\ \bibinfo {author} {\bibfnamefont {G.~E.}\ \bibnamefont {Scuseria}},\
  }\href@noop {} {\bibfield  {journal} {\bibinfo  {journal} {J. Chem. Phys.}\
  }\textbf {\bibinfo {volume} {146}},\ \bibinfo {pages} {184105} (\bibinfo
  {year} {2017}{\natexlab{b}})}\BibitemShut {NoStop}%
\bibitem [{\citenamefont {Degroote}\ \emph {et~al.}(2016)\citenamefont
  {Degroote}, \citenamefont {Henderson}, \citenamefont {Zhao}, \citenamefont
  {Dukelsky},\ and\ \citenamefont {Scuseria}}]{Degroote16}%
  \BibitemOpen
  \bibfield  {author} {\bibinfo {author} {\bibfnamefont {M.}~\bibnamefont
  {Degroote}}, \bibinfo {author} {\bibfnamefont {T.~M.}\ \bibnamefont
  {Henderson}}, \bibinfo {author} {\bibfnamefont {J.}~\bibnamefont {Zhao}},
  \bibinfo {author} {\bibfnamefont {J.}~\bibnamefont {Dukelsky}}, \ and\
  \bibinfo {author} {\bibfnamefont {G.~E.}\ \bibnamefont {Scuseria}},\
  }\href@noop {} {\bibfield  {journal} {\bibinfo  {journal} {Phys. Rev. B}\
  }\textbf {\bibinfo {volume} {93}},\ \bibinfo {pages} {125124} (\bibinfo
  {year} {2016})}\BibitemShut {NoStop}%
\end{thebibliography}
\end{document}